\documentclass[preprint2]{aastex}
\usepackage[colorlinks]{hyperref}
\usepackage{amsmath}
\usepackage{algorithm}
\usepackage{algorithmic}
\usepackage{mathtools}
\usepackage{acronym}

\acrodef{BH}[BH]{black hole}
\acrodef{CBC}[CBC]{compact binary coalescence}
\acrodef{EM}[EM]{electromagnetic}
\acrodef{DFT}[DFT]{discrete Fourier transform}
\acrodef{FFT}[FFT]{fast Fourier transform}
\acrodef{FIR}[FIR]{finite impulse response}
\acrodef{FOV}[FOV]{field of view}
\acrodefplural{FOV}[FOVs]{fields of view}
\acrodef{GRB}[GRB]{$\gamma$\nobreakdashes-ray burst}
\acrodef{GW}[GW]{gravitational\nobreakdashes-wave}
\acrodef{LIGO}[LIGO]{Laser Interferometer \acs{GW} Observatory}
\acrodef{MCMC}[MCMC]{Markov chain Monte Carlo}
\acrodef{NS}[NS]{neutron star}
\acrodef{SSC}[SSC]{synchrotron self\nobreakdashes-Compton}
\acrodef{CMB}[CMB]{cosmic microwave background}
\acrodef{ROTSE}[ROTSE]{Robotic Optical Transient Search}
\acrodef{TAROT}[TAROT]{T\'{e}lescopes \`{a} Action Rapide pour les Objets Transitoires}
\acrodef{CCD}[CCD]{charge coupled device}
\acrodef{QUEST}[QUEST]{Quasar Equatorial Survey Team}
\acrodef{PTF}[PTF]{Palomar Transient Factory}
\acrodef{LOFAR}[LOFAR]{Low Frequency Array}
\acrodef{UFFO}[UFFO]{Ultra Fast Flash Observatory}
\acrodef{BAT}[BAT]{Burst Alert Telescope}
\acrodef{SMT}[SMT]{Slewing Mirror Telescope}
\acrodef{IACT}[IACT]{imaging atmospheric \v{C}erenkov telescope}
\acrodef{VHE}[VHE]{very high energy}
\acrodef{UHE}[UHE]{ultra high energy}
\acrodef{HETE}[HETE]{High Energy Transient Explorer}
\acrodef{CGRO}[CGRO]{Compton Gamma Ray Observatory}
\acrodef{XRT}[XRT]{X\nobreakdashes-ray Telescope}
\acrodef{SNR}[SNR]{signal\nobreakdashes-to\nobreakdashes-noise ratio}
\acrodef{aLIGO}[aLIGO]{Advanced \acs{LIGO}}
\acrodef{AdVirgo}[AdVirgo]{Advanced Virgo}
\acrodef{BAYESTAR}[BAYESTAR]{BAYEsian optimal Search for Transients with Autonomous and Robotic telescopes}
\acrodef{DASWG}[DASWG]{Data Analysis Software Working Group}
\acrodef{NSF}[NSF]{National Science Foundation}
\acrodef{REU}[REU]{Research Experiences for Undergraduates}
\acrodef{LSST}[LSST]{Large Synoptic Survey Telescope}
\acrodef{SHGRB}[SHGRB]{short, hard \acl{GRB}}

\newcommand{\npix}{{n_\mathrm{pix}}}
\DeclareMathOperator{\re}{Re}
\DeclareMathOperator{\im}{Im}

\begin{document}

\title{Optimizing optical follow\nobreakdashes-up of gravitational\nobreakdashes-wave candidates}

\author{Leo Singer\email{leo.singer@ligo.org}Larry Price\email{larryp@caltech.edu}}
\affil{LIGO Laboratory, California Institute of Technology}
\affil{MC 100-36, 1200 E. California Blvd., Pasadena, CA 91101, USA}
\author{Antony Speranza}
\email{asperanz@mit.edu}
\affil{Department of Physics, Massachusetts Institute of Technology}
\affil{4-304, 77 Massachusetts Ave., Cambridge, MA 02139, USA}

\shorttitle{Optimizing optical follow\nobreakdashes-up}
\shortauthors{Price, Singer, and Speranza}


\begin{abstract}
Observations with interferometric gravitational\nobreakdashes-wave detectors result in probability sky maps that are multimodal and spread over $\gtrsim$10\nobreakdashes--100~deg$^2$.  We present a scheme for maximizing the probability of imaging optical counterparts to gravitational\nobreakdashes-wave transients given limited observing resources.  Our framework is capable of coordinating many
telescopes with different \aclp{FOV} and limiting magnitudes.  We present a case study comparing three different planning algorithms.  We find that, with the network of telescopes that was used in the most recent joint \acs{LIGO}--Virgo science run, a relatively straightforward coordinated approach doubles the detection efficiency relative to each telescope observing independently.
\end{abstract}

\section{Introduction}

Among the most promising sources for coincident \ac{EM} and \ac{GW} emission are \ac{CBC} events in which two \acp{NS} or a \ac{NS} and a \ac{BH} inspiral and merge.  The \ac{NS} may be torn apart before merger, powering \ac{EM} emission at a variety of timescales and wavelengths ranging from seconds to months in $\gamma$\nobreakdashes- and X\nobreakdashes-ray to radio, respectively~\citep{0034-4885-69-8-R01, 2007PhR...442..166N, MetzgerBerger:2012}.  It is suspected that \acp{CBC} are also the progenitors \citep{0004-637X-648-2-1110, 1367-2630-9-1-017, MNL2:MNL20421} of some or all \acp{SHGRB}.  Plausible \ac{CBC} event rates suggest that \acl{aLIGO}~\citep[\acs{aLIGO};][]{aLIGO} and \acl{AdVirgo}~\citep[\acs{AdVirgo};][]{AdVirgo} could detect about 40 \ac{NS}\nobreakdashes--\ac{NS} and 10 \ac{NS}\nobreakdashes--\ac{BH} events per year of observation time \citep{rates}.  Aside from shedding light
on the true progenitors of \acp{SHGRB}~\citep{sylvestre, lee}, coincident
detections of \ac{EM} and \ac{GW} emission could also be utilized as ``standard
sirens'' to provide a precision measurement of the Hubble
constant~\citep{schutz2001, Nissanke:2010}.

In addition to \acp{CBC}, there are other possible sources of coincident
\ac{EM} and \ac{GW} emission.  Examples include type II supernovae, which are much more common than \acp{CBC} at $\lesssim$ 1 event per 20 years in the Milky Way alone, but whose \ac{GW} counterparts would be detectable by \ac{aLIGO} only within the local group~\citep{ott}.  Soft $\gamma$ repeaters \citep{MNR:MNR4756, PhysRevLett.101.211102} and anomalous X\nobreakdashes-ray pulsars \citep{2041-8205-734-2-L35} could produce \acp{GW}, and some models predict \ac{GW} emission comparable to the best upper limits \citep{PhysRevD.83.104014}.  Neutron stars recovering from pulsar glitches might produce \ac{GW} transients on timescales of days to weeks that could be detectable at $\lesssim$10~kpc \citep{MNR:MNR17416}.  More speculative sources of \ac{GW} include cosmic string cusps~\citep{cscprl,cscgw}, which if they exist might also power \acsp{GRB} \citep{1987ApJ...316L..49B, PhysRevD.64.043004, PhysRevLett.106.259001} and radio bursts \citep{cscem}.

The advent of deep, wide\nobreakdashes-field optical survey telescopes and rapid response robotic telescopes is paving the way for time\nobreakdashes-domain astronomy.  The simultaneous construction of advanced
\ac{GW} detectors creates the enticing possibility of a coincident
detection of a \ac{GW} transient and an associated \ac{EM} counterpart.  Efforts in this direction were recently made in \acs{LIGO}'s sixth and Virgo's second and third science runs, which saw the introduction of a program~\citep{loocup:2008, loocup:2012, CBCLowLatency, SwiftFollowup, OpticalImageAnalysis} whereby potential \ac{GW} transient alerts were sent to X\nobreakdashes-ray, optical, and radio facilities for follow\nobreakdashes-up.  In the context of space-based \citep{kocsis:2008} and advanced ground-based \citep{fairhurst:2009, lloid:2012, dietz:2012} \ac{GW} detectors, there is even discussion of using \acp{GW} to trigger early\nobreakdashes-warning follow\nobreakdashes-up observations while the inspiral is still ongoing.

The great disparity in the character of optical and \ac{GW} observations makes \ac{GW}\nobreakdashes-triggered detection of optical counterparts challenging.  Although there are now several operating ultra\nobreakdashes-wide\nobreakdashes-field, but relatively shallow, transient monitoring telescopes such as Pi of the Sky, only a handful of operating or planned deep survey telescopes have \acp{FOV} that are comparable to the error regions of $\sim$10\nobreakdashes--100~deg$^2$ \citep{fairhurst:2009, Wen:2010, Nissanke:2011, fairhurst:2011} that will be typical of \ac{GW} sky maps in the advanced detector era.  By contrast, a \ac{GW} detector is both blessed and cursed by its nature as an all\nobreakdashes-sky instrument: it has coarse directional sensitivity only due to its geometry\nobreakdashes-defined quadrupole antenna pattern.

In this paper we lay a framework for planning optimal \ac{EM} follow\nobreakdashes-up of \ac{GW} candidates.  Our paradigm focuses on tiling the
\ac{GW} sky map as efficiently as possible by coordinating a number of
telescopes with dissimilar \acp{FOV}.  Although our formalism supports it, we do not yet account for seeing, different limiting magnitudes, different times at which observations are made, or multiple observations with each telescope.
As preliminaries, we describe the key features of ground\nobreakdashes-based interferometric
\ac{GW} detectors as contrasted with optical telescopes.  We demonstrate that the optimal pointing of a single telescope can be expressed as a convolution\nobreakdashes-like integral on the unit sphere, which is amenable to spherical harmonic analysis.  For multiple telescopes, we provide a selection of observation planning algorithms that take into account the particular geometry of each telescope's image plane as well as local observing conditions (in our present implementation, just the position of the sun and the Earth).   We have developed a parallelized observation planning code in Python/C/C++ called
\ac{BAYESTAR}%
\footnote{A pun on the Cylon battleships in the
American television series Battlestar Galactica.  The defining characteristic of
the Cylons is that they repeatedly defeat humanity by using their superhuman
information\nobreakdashes-gathering ability to coordinate overwhelming forces.  The name
also suggests that, like the Cylons, robotic transient telescopes may some day
rise against us humans.}%
\footnote{We do not like to mention the final `T' in the
acronym, because then it would be called BAYES\nobreakdashes-TART, which would sound
ridiculous.}%
, which can map out follow\nobreakdashes-up telescope pointings with low latency in the way described in this work.  We show that with the telescopes that were available for
\ac{EM} follow\nobreakdashes-up in the last \acs{LIGO}\nobreakdashes--Virgo science run, if each telescope performs just one pointing,
the imaging efficiency is doubled by pointing all of the telescopes in coordination
rather than independently.  Finally, we propose an expanded
simulation campaign and discuss some practical issues for \ac{EM} follow\nobreakdashes-up in the advanced
\ac{GW} detector era.

\section{\ac{GW} detectors and optical telescopes}

In this section, we review the observational capabilities of \ac{GW} detectors as contrasted with optical telescopes.

\ac{GW} detectors such as \acs{LIGO}~\citep{iLIGO} and Virgo~\citep{iVirgo} are Fabry\nobreakdashes--Perot Michelson laser interferometers that sense \ac{GW}\nobreakdashes-induced strain through differences in light travel time in two arms.  The only directional sensitivity of a single \ac{GW} detector is due to its antenna pattern~\citep[see, for example,][]{lrr-2009-2}.  As predicted in general relativity, \acp{GW} are transverse and come in two orthogonal polarizations, the so\nobreakdashes-called `$+$' and `$\times$' modes.  An interferometer with perpendicular arms is most sensitive to a `$+$' polarized source at the zenith (or nadir), with the sensitivity going to zero at any azimuth that is 45$^\circ$ from either arm.  The sensitivity to the `$\times$' mode is also greatest at the zenith, but vanishes at any azimuth that is at a right angle to either arm.  It also vanishes everywhere in the plane of the detector.

A signal's time delay on arrival at two \ac{GW} detectors constrains the position of the \ac{GW} source to an annulus on the sky.  The antenna patterns of the two detectors impose constraints that disfavor parts of the annulus~\citep{Raymound:2009p114007}.  Time delay measurements from three gravitational wave detectors constrain a source to two regions that are located at mirroring points on either side of the plane formed by the detectors.  The mirror degeneracy is, again, partly broken by the antenna patterns.  Networks of four or more detectors are sufficient to determine the location of the source to within a single patch on the sky.  The network consisting of \ac{AdVirgo} and the two \ac{aLIGO} observatories will be able to constrain sources to areas of 10\nobreakdashes--100~deg$^2$ with 90\% confidence, varying across the sky as a result of the antenna patterns~\citep{Wen:2010, Nissanke:2011, fairhurst:2011, klimenko:2011}.  Adding the planned Japanese detector KAGRA \citep[formerly LCGT;][]{LCGT} or the proposed \acs{LIGO}\nobreakdashes--India detector would narrow the confidence region~\citep{fairhurst:2011,0264-9381-28-12-125023}.

Optical telescopes are markedly different from gravitational wave detectors in that they have directional precision of arcseconds or better.  This is at the expense of observing only a small fraction of the sky at one time.  Table~\ref{tab:telescope-list} shows a list of telescopes that participated in the \acs{LIGO}\nobreakdashes--Virgo \ac{EM} follow\nobreakdashes-up program.  \acp{FOV} vary from scarcely 0.01~deg$^2$ with relatively deep instruments such as Liverpool Telescope at a limiting 21~mag; to a few deg$^2$ with small robotic transient follow\nobreakdashes-up telescopes such as \acs{ROTSE} at 17~mag; to many deg$^2$ with survey instruments such as Skymapper and \ac{PTF}, at $\gtrsim$20~mag.  The largest \acp{FOV} of $\gtrsim$100~deg$^2$ belong to transient monitoring instruments such as Pi of the Sky at $\gtrsim$12~mag.  The eagerly awaited \ac{LSST} will have an even larger \ac{FOV} than \ac{PTF}, but with a limiting magnitude of 24.5 in a similar exposure time.  Slew times can be as brief as seconds for small robotic telescopes like \acs{ROTSE} and \acs{TAROT}, or minutes and longer for larger instruments or instruments that are less optimized for response time.  Typical total exposure times range from $\sim$10~s to a few minutes per pointing.

\begin{deluxetable}{r@{}lr@{$\times$}ll@{ in }llr@{, }l}
\tablecaption{Telescopes Employed in Joint \acs{LIGO}\nobreakdashes--Virgo Science Run\label{tab:telescope-list}}
\tablehead{& & \multicolumn{2}{c}{} & \multicolumn{2}{l}{limiting} & slew & \multicolumn{2}{c}{} \\
site & & \multicolumn{2}{c}{field of view} & \multicolumn{2}{l}{magnitude} & time (s) & \multicolumn{2}{c}{geographic location}}
\startdata
Liverpool&$^\text{ab\tablenotemark{1}}$ & 0.077$^\circ$ & 0.077$^\circ$ & 21 & 300 s & 30 & $28^\circ 45' 44.8'' \textrm{ N}$ & $ 17^\circ 52' 45.2'' \textrm{ W}$ \\
Zadko&$^\text{c}$ & 1.4$^\circ$ & 1.4$^\circ$ & 21 & 180 s & 20 & $31^\circ 21' 24'' \textrm{ S}$ & $ 155^\circ 42' 49'' \textrm{ E}$ \\
ROTSE III-a&$^\text{d}$ & 1.85$^\circ$ & 1.85$^\circ$ & 17 & 5 s & 4 & $31^\circ 16' 24.1'' \textrm{ S}$ & $ 149^\circ 3' 40.3'' \textrm{ E}$ \\
ROTSE III-b&$^\text{d}$ & 1.85$^\circ$ & 1.85$^\circ$ & 17 & 5 s & 4 & $23^\circ 16' 18'' \textrm{ S}$ & $ 16^\circ 30' 00'' \textrm{ E}$ \\
ROTSE III-c&$^\text{d}$ & 1.85$^\circ$ & 1.85$^\circ$ & 17 & 5 s & 4 & $36^\circ 49' 30'' \textrm{ N}$ & $ 30^\circ 20' 0'' \textrm{ E}$ \\
ROTSE III-d&$^\text{d}$ & 1.85$^\circ$ & 1.85$^\circ$ & 17 & 5 s & 4 & $30^\circ 40' 17.7'' \textrm{ N}$ & $ 104^\circ 1' 20.1'' \textrm{ W}$ \\
TAROT&$^\text{ef}$ & 1.86$^\circ$ & 1.86$^\circ$ & 17 & 10 s & 1.5 & $43^\circ 45' 8'' \textrm{ N}$ & $ 6^\circ 55' 26'' \textrm{ E}$ \\
TAROT-S&$^\text{ef}$ & 1.86$^\circ$ & 1.86$^\circ$ & 17 & 10 s & 1.5 & $29^\circ 15' 39'' \textrm{ S}$ & $ 70^\circ 43' 56'' \textrm{ W}$ \\
Skymapper&$^\text{g}$ & 2.373$^\circ$ & 2.395$^\circ$ & 21.6 & 110 s &  & $31^\circ 16' 24'' \textrm{ S}$ & $ 149^\circ 3' 52'' \textrm{ E}$ \\
PTF&$^\text{h}$ & 3.5$^\circ$ & 2.31$^\circ$ & 20.6 & 60 s &  & $33^\circ 21' 21'' \textrm{ N}$ & $ 116^\circ 51' 50'' \textrm{ W}$ \\
QUEST&$^\text{i}$ & 3.6$^\circ$ & 4.6$^\circ$ & 21 & 140 s &  & $33^\circ 21' 21'' \textrm{ N}$ & $ 116^\circ 51' 50'' \textrm{ W}$ \\
Pi of the Sky South&$^\text{jk}$ & 20$^\circ$ & 20$^\circ$ & 12 & 10 s & 60 & $22^\circ 57' 12'' \textrm{ S}$ & $ 68^\circ 10' 48'' \textrm{ W}$ \\
Pi of the Sky North&$^\text{jk}$ & 40$^\circ$ & 40$^\circ$ & 12 & 10 s & 40 & $37^\circ 6' 14'' \textrm{ N}$ & $ 6^\circ 44' 3'' \textrm{ W}$ \\
\enddata
\tablecomments{\ac{FOV}, limiting magnitude, slew time, and geographic location of ground-based robotic optical telescopes that participated in the \acs{LIGO}\nobreakdashes--Virgo \ac{EM} follow\nobreakdashes-up program.  Except where otherwise indicated, we quote the shortest exposure for which a limiting magnitude is available in the literature.}
\tablenotetext{1}{\acs{LIGO}'s request for Liverpool Telescope time calls for a 300~s exposure based on a limiting magnitude of 21~\citep{white:2012}.}
\tablerefs{ $^\text{a}${\citet{liverpool:2004}};  $^\text{b}${\citet{liverpool:2006}};  $^\text{c}${\citet{zadko:2010}};  $^\text{d}${\citet{rotse:2003}};  $^\text{e}${\citet{tarot:2003}};  $^\text{f}${\citet{ tarot:2008}};  $^\text{g}${\citet{skymapper:2007}};  $^\text{h}${\citet{ptf:2009}};  $^\text{i}${\citet{quest:2007}};  $^\text{j}${\citet{pi:2010}};  $^\text{k}${\citet{pi:2011}};  }
\end{deluxetable}

\section{Single telescope case}

In this section, we introduce the optimization problem of pointing a single telescope to maximize the probability of imaging the true, but unknown, location of the \ac{GW} source.  Then, we describe two numerical procedures to find the optimum.

For \ac{GW} detectors, the observables are strain time series consisting of noise plus the astrophysical \ac{GW} signal.  The \ac{GW} signal from a \ac{CBC} source is determined by extrinsic parameters consisting of right ascension $\phi$, declination $(\pi / 2 - \theta)$, luminosity distance $D_L$, inclination angle $\iota$, coalescence time $t$, polarization angle $\psi$, and orbital phase at coalescence $\varphi$; and intrinsic parameters including the component masses and spins.  One may form the posterior distribution of all of these parameters:
$$
	p(\theta, \phi, D_L, \iota, t, \psi, \varphi, m_1, m_2, \dots | \textsc{gw}).
$$
The proposition $\textsc{gw}$ represents the entire \ac{GW} observation: the strain in each \ac{GW} detector, the geographic locations of the detectors, and their antenna patterns.  If one fixes or marginalizes over all parameters except for the position on the sky, the spherical polar coordinates $\boldsymbol\omega=(\theta, \phi)$, then this reduces to what we term the \ac{GW} sky map,
$$
	p(\boldsymbol\omega | \textsc{gw}) = s(\boldsymbol\omega).
$$
This quantity can be evaluated from time delays and antenna factors~\citep{CBCLowLatency} or from the \ac{GW} strain itself using analytic methods~\citep{Searle:2008,Searle:2009} or \ac{MCMC} techniques~\citep{PhysRevD.58.082001, 0264-9381-25-18-184011, PhysRevD.81.062003, Nissanke:2011}.

Consider the $i$th of $N$ telescopes labeled $1 \dots N$.  Let the proposition $\textsc{em}_i$ denote the detection of an associated optical transient with telescope $i$.  Let $\mathbf C_i$ represent the configuration of telescope $i$, which consists of a pointing $\boldsymbol\gamma_i = (\theta_i, \phi_i)$, nominal limiting magnitude, exposure start and end times, as well as uncontrollable factors such as seeing and interference from the horizon, sun, moon, and the galactic disk.  If we knew the position $\boldsymbol\omega$ of the source, we could evaluate the probability of detecting an \ac{EM} counterpart at $\boldsymbol\omega$ with telescope $i$,
$$
	p(\textsc{em}_i | \mathbf \mathbf C_i, \boldsymbol\omega).
$$
For most of this paper we shall restrict ourselves to the dependence on the pointing of the telescope, so that the detection probability factors into a function $b_i(\boldsymbol\omega)$ describing the telescope's \ac{FOV} in a home orientation (i.e., for equatorial mounts, the celestial North pole) and a function $v_i(\boldsymbol\omega)$ that describes environmental factors including the nighttime/twilight terminator and the horizon.  In principal, $v_i(\boldsymbol\omega)$ could also include other observational factors such as moonlight and current seeing.  The detection probability becomes:
\begin{equation}
	p(\textsc{em}_i | \boldsymbol\gamma_i, \boldsymbol\omega) = b_i\left(R^{-1}[\boldsymbol\gamma_i]\, \boldsymbol\omega\right) v_i(\boldsymbol\omega).
\end{equation}
A simple rectangular \ac{FOV} with angular width $w$ and height $h$ is described by
\begin{equation}
	b_i(\theta, \phi) =
		\begin{cases}
			1 &
				\begin{split}
					\text{if } |\sin \theta \cos \phi| &\leq \sin {w/2}, \\
					|\sin \theta \sin \phi| &\leq \sin {h/2}, \\
					\text{and } \theta &< \pi / 2;
				\end{split} \\
			0 & \text{otherwise.}
		\end{cases}
\end{equation}
Much more complicated \acp{FOV} are commonplace because a real telescope may have a focal plane consisting of a \acs{CCD} mosaic with a filling factor that is much less than unity; dead \acsp{CCD} or \acsp{CCD} that contain defects; or \acsp{CCD} that are vignetted or clipped by the optics.

$R[\boldsymbol\gamma_i]$ denotes an active rotation operation about the $z$\nobreakdashes-axis by $\phi_i$ and about the $y$\nobreakdashes-axis by $\theta_i$, so $R^{-1}[\boldsymbol\gamma_i]$ is the passive rotation such that $b_i$ evaluated at $\left(R^{-1}[\boldsymbol\gamma_i]\, \boldsymbol\omega\right)$ represents the \ac{FOV} centered on the sky location $\boldsymbol\gamma_i = (\theta_i, \phi_i)$.  This rotation matrix is described by two, not three, angles because mounts for ground\nobreakdashes-based telescopes usually have two degrees of freedom (for example, right ascension and declination or azimuth and altitude).

This particular model incorporates the assumption that a telescope always detects an optical transient when the true location of the \ac{GW} source is inside the \ac{FOV}.  We can identify this with the probability of \textit{imaging} the source, which for most of this work we shall use as a proxy for the ultimately desirable probability of \textit{detection}.

The optimal configuration $\mathbf C_i^\mathrm{*}$ of telescope $i$ maximizes the probability of detection of an electromagnetic counterpart given all gravitational wave observations, marginalized over the unknown source location.  To wit,
\begin{equation}
	\mathbf C_i^\mathrm{*} = \arg \max_{\mathbf C_i} \int
		p(\textsc{em}_i | \mathbf C_i, \boldsymbol\omega) \,
		p(\boldsymbol\omega | \textsc{gw}) \, \mathrm{d}\boldsymbol{\Omega},
\end{equation}
where the integral is over $\boldsymbol\omega$ and $\mathrm{d}\boldsymbol{\Omega}$ represents the surface element on the unit sphere.  In the simplified scenario in which the pointing orientation is the only free parameter, the optimal pointing is given by
\begin{align}
	\label{eq:optimal-pointing-single-telescope}
	\boldsymbol\gamma_i^\mathrm{*} &= \arg \max_{\boldsymbol\gamma_i} \int
		p(\textsc{em}_i | \boldsymbol\gamma_i, \boldsymbol\omega) \,
		p(\boldsymbol\omega | \textsc{gw}) \, \mathrm{d}\boldsymbol{\Omega} \nonumber \\
		&= \arg \max_{\boldsymbol\gamma_i} \int
			s(\boldsymbol\omega) \, b_i\left(R^{-1}[\boldsymbol\gamma_i]\, \boldsymbol\omega\right) \, v_i(\boldsymbol\omega) \, \mathrm{d}\boldsymbol{\Omega}.
\end{align}

The integral in Equation~(\ref{eq:optimal-pointing-single-telescope}) resembles a convolution integral, but over rotations instead of over the real number line.  Just as \ac{FFT} methods are used to efficiently evaluate convolutions, fast spherical harmonic methods may be used to evaluate Equation~(\ref{eq:optimal-pointing-single-telescope}) efficiently.  In the Appendix, we describe in detail two algorithms for solving the single telescope problem: a ``direct'' or spatial algorithm and a ``multipole'' algorithm based on the method for fast convolution on the sphere of \citet{wandelt}.  Both algorithms exploit special properties of the HEALPix\footnote{\url{http://healpix.jpl.nasa.gov/}}~\citep{healpix} projection.  Either of these algorithms may be used as a component of the planning of observations with multiple telescopes, which we discuss in the next section.

\section{Multiple telescope case}

In order to address the problem of pointing multiple telescopes, we must first specify a figure of merit.  One example is the probability of observing an electromagnetic counterpart in any (one or more) telescope,
\begin{equation}
	\label{eq:p_any_definition}
	p_\mathrm{any} \equiv p(
		\textsc{em}_1 \cup \textsc{em}_2 \cup \dots
		\cup \textsc{em}_N | \boldsymbol\gamma_1, \boldsymbol\gamma_2, \dots, \boldsymbol\gamma_N, \textsc{gw}).
\end{equation}
This only considers the case in which each telescope is asked to point once.  Some telescopes will take a sequence of images at arbitrary sky locations.  Others are constrained to follow a particular path on the sky, for instance, visiting reference pointings in a specified order.  In the former case, we could trivially interpret $\textsc{em}_1 \dots \textsc{em}_N$ as optical detections in the \emph{observations} $1 \dots N$ at times $t_1 \dots t_N$ respectively, where there are $N$ observations but potentially $\ll N$ telescopes.  In the latter case, we could define a composite \ac{FOV} that describes the probability of imaging the source in any of the pointings throughout the entire observation.  A more elaborate objective function would also include a penalty on the number of pointings in order to economize the limited observational resources.  For the remainder of this paper, we will focus on $p_\mathrm{any}$%
\footnote{
Alternatively, we could demand not one, but several, \ac{EM} detections to secure the discovery of a bona fide optical transient.  Combinatorial considerations give the probability of detection in at least $K$ out of $N$ observations,
\begin{align}
p(\text{at least } &K \text{ detections}\, | \,\textsc{gw}) = \nonumber\\
\int \Bigg[
& (-1)^{K + 1} \binom{0}{K-1} \sum_i p(\textsc{em}_i | \gamma_i, \boldsymbol\omega) \nonumber\\
+ & (-1)^{K + 2} \binom{1}{K-1} \sum_{i \neq j} p(\textsc{em}_i \cap \textsc{em}_j | \gamma_i, \gamma_j, \boldsymbol\omega) \nonumber\\
+ & (-1)^{K + 3} \binom{2}{K-1} \sum_{\mathclap{i \neq j \neq k}} p(\textsc{em}_i \cap \textsc{em}_j \cap \textsc{em}_k | \gamma_i, \gamma_j, \gamma_k, \boldsymbol\omega) \nonumber\\
+ & \cdots \nonumber\\
+ & (-1)^{K + N} \binom{N-1}{K-1} \; p(\textsc{em}_1 \cap \textsc{em}_N | \gamma_1, \dots, \gamma_N, \boldsymbol\omega)\Bigg] \nonumber\\
& p(\boldsymbol\omega | \textsc{gw}) \, \mathrm{d}\boldsymbol{\Omega}.
\end{align}
}%
.

De Morgan's law tells us that Equation~(\ref{eq:p_any_definition}) is equivalent to the complement of the probability of non\nobreakdashes-detection with every telescope,
\begin{align}\label{eq:p_any}
	p_\mathrm{any} &= 1 - \int \left( \prod_i \left[1 - p(\textsc{em}_i | \boldsymbol\gamma_i, \boldsymbol\omega)\right] \right) p(\boldsymbol\omega | \textsc{gw}) \, \mathrm{d}\boldsymbol{\Omega} \nonumber\\
		&= 1 - \int
		\left[ 1 - b_1 \left(R^{-1}[\boldsymbol\gamma_1]\, \boldsymbol\omega\right) v_1(\boldsymbol\omega) \right] \nonumber\\
		&\phantom{\,= 1 - \int} \left[ 1 - b_2 \left(R^{-1}[\boldsymbol\gamma_2]\, \boldsymbol\omega\right) v_2(\boldsymbol\omega) \right] \nonumber\\
		&\phantom{\,= 1 - \int}\mathllap{\cdots}
		\left[ 1 - b_N \left(R^{-1}[\boldsymbol\gamma_N]\, \boldsymbol\omega\right) v_N(\boldsymbol\omega) \right]
		s(\boldsymbol\omega) \, \mathrm{d}\boldsymbol{\Omega}.
\end{align}
The optimal set of pointings maximizes $p_\mathrm{any}$,
\begin{equation}
	(\boldsymbol\gamma_1^*, \boldsymbol\gamma_2^*, \dots, \boldsymbol\gamma_N^*) \equiv \arg \max_{\mathclap{\boldsymbol\gamma_1, \boldsymbol\gamma_2, \dots, \boldsymbol\gamma_N}} \; p_\mathrm{any}.
\end{equation}

\subsection{Planning algorithms}

We considered three different multiple telescope algorithms, increasing in sophistication and accuracy but also increasing in computational overhead.  After describing each algorithm in detail, we exhibit implementations that use the positional astronomy library NOVAS%
\footnote{\url{http://aa.usno.navy.mil/software/novas/novas_info.php}}
\citep{NOVAS} to take into account rudimentary sun and horizon interference assuming observers at sea level on an atmosphereless, perfectly spherical Earth.  We assume that all observations occur simultaneously at the time of the \ac{GW} detection.

\subsubsection{Noncoordinated}

The \emph{noncoordinated} planner models the situation in which each telescope operator independently decides to observe that patch of the sky that maximizes his or her own odds of detecting a counterpart.  Each single telescope problem is solved independently, regardless of the configurations of any of the other telescopes.

The \emph{noncoordinated} method has the disadvantage that it will generally result in all of the telescopes pointing at the same region of the sky, perhaps failing to image all of the significant regions of the \ac{GW} sky map.  However, if the telescope network's coverage is exceedingly poor, such that the nonzero regions of the sky map are within view of only $\sim 1$ telescope at a time, then this will not be a significant drawback.  The \emph{noncoordinated} method also has the advantage that it is exceedingly simple to implement.  For these reasons, the \emph{noncoordinated} planner may represent a reasonable, but sub\nobreakdashes-optimal, observational strategy.

\subsubsection{Greedy}

\begin{figure}
	\begin{center}
		\includegraphics{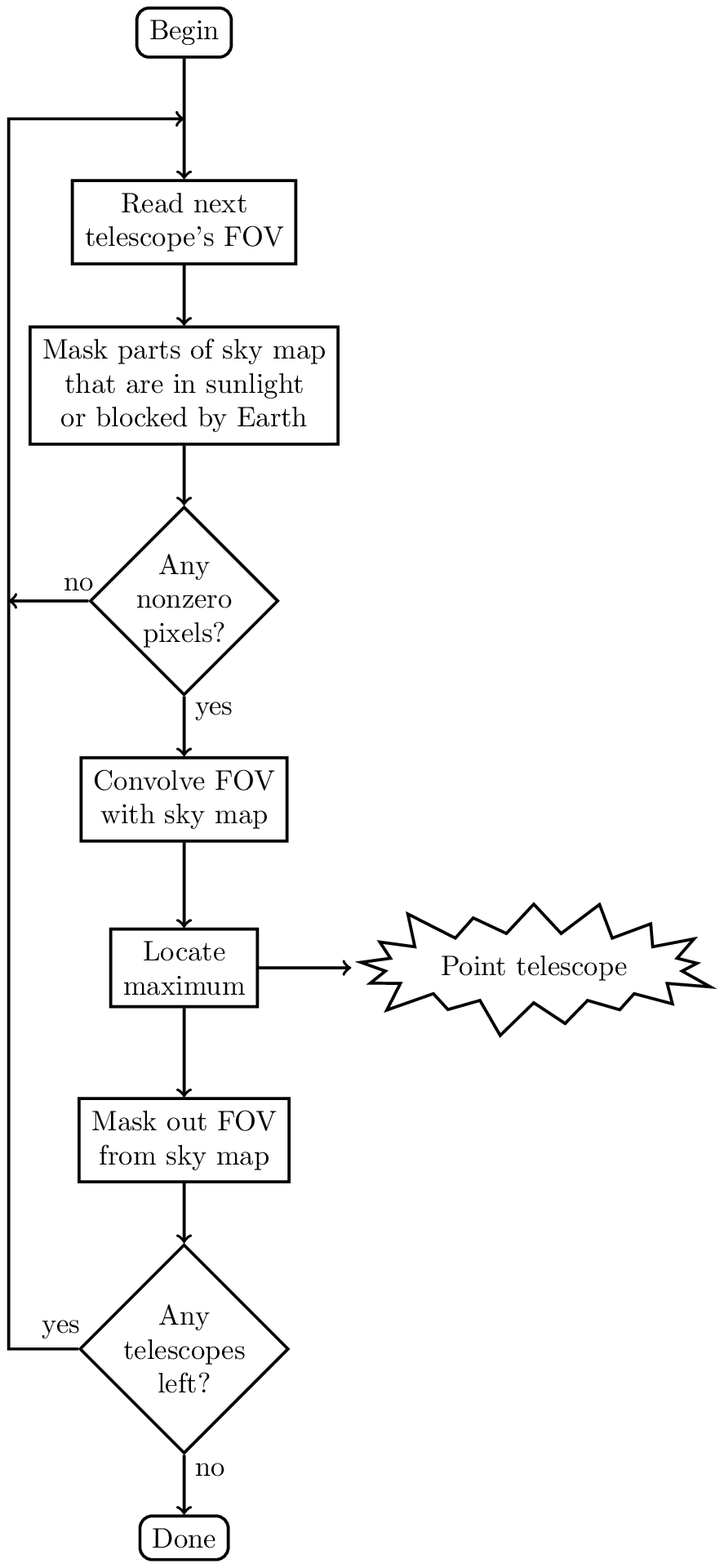}
	\end{center}
	\caption{\label{fig:greedy}Flow chart depicting greedy algorithm.}
\end{figure}

The \emph{greedy} planner is slightly more sophisticated than the \emph{noncoordinated} method.  Suppose that pointings have been decided upon for telescopes $1, 2, \dots, i$.  Then the configuration of telescope $i + 1$ is chosen such that it maximizes the imaging probability when all of the previous pointings $1 \dots i$ are held fixed.  This procedure is illustrated in Figure~\ref{fig:greedy}.  It is not guaranteed to find an absolute maximum, but we shall see that in practice it does quite well.

This algorithm leaves a choice of the order in which to point the telescopes.  This is analogous to the problem of packing a number of stones of different sizes into a bucket.  The intuitive answer is to insert the largest rocks first and then use the pebbles to fill in the gaps.  Likewise, we find that we obtain the best detection efficiency when we sort the telescopes by decreasing \ac{FOV} area because the telescope with the largest \ac{FOV} makes the greatest progress toward imaging the entire support of the \ac{GW} sky map.  We refer to the greedy algorithm with this prescription as the \emph{greedy\nobreakdashes-sorted} planner.  We expect that the \emph{greedy\nobreakdashes-sorted} planner will always outperform the \emph{noncoordinated} planner by a factor that depends on the \ac{SNR} of the \ac{GW} signal and the particular network of telescopes.

\subsubsection{Simulated annealing}

The \emph{anneal} planner uses the simulated annealing code provided by the SciPy library%
\footnote{\url{http://docs.scipy.org/doc/scipy/reference/generated/scipy.optimize.anneal.html}}
to extremize $p_\mathrm{any}$.  Starting from an initially proposed observing plan, this procedure randomly perturbs all of the telescopes' pointings simultaneously.  If the imaging probability has improved, the perturbed plan is adopted.  The perturbations become smaller and smaller with each iteration; the perturbation distribution as a function of time is called a \emph{cooling schedule}.  We used a modified version of the ``very fast'' cooling schedule described by \citet{Ingber1989967} with an initial ``temperature'' $T_0 = 10^{-4}$.  Since \citeauthor{Ingber1989967}'s cooling schedule uses rectilinear coordinates, we modified it by perturbing the Cartesian coordinates of the center of each telescope's \ac{FOV} and then normalizing to unit length.

Simulated annealing will nearly always find the globally optimal observing plan in a finite, but possibly large, number of iterations.  With a particular fixed number of iterations, we expect that it will usually outperform not only the \emph{noncoordinated} planner but also the \emph{greedy\nobreakdashes-sorted} planner.  However, the computational overhead of evaluating thousands of observing plans in sequence may be prohibitive if there is a time constraint---if, for example, it is important to begin taking images as soon as possible after the \ac{GW} event.

\subsection{Case study}

\begin{figure*}
	\begin{center}
		\begin{minipage}[b]{0.3\textwidth}
			\begin{center}
				\includegraphics{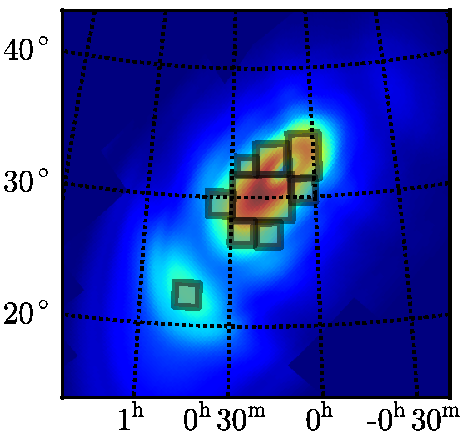}  \\
				(a) Detail ``a''
			\end{center}
		\end{minipage}%
		\begin{minipage}[b]{0.3\textwidth}
			\begin{center}
				\includegraphics{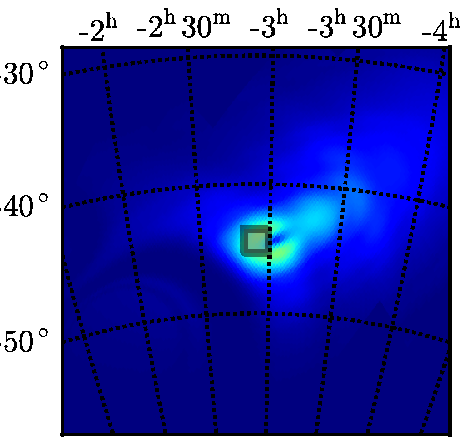} \\
				(b) Detail ``b''
			\end{center}
		\end{minipage}%
		\begin{minipage}[b]{0.4\textwidth}
			\begin{center}
				\begin{tabular}{lll}
\multicolumn{3}{l}{GPS time = 894384569.0} \\
\tableline\tableline
site & RA & Dec \\
\tableline
QUEST & $00^\mathrm{h} 19^\mathrm{m} 28^\mathrm{s}$ & $030^\circ 05'$ \\
PTF & $00^\mathrm{h} 03^\mathrm{m} 31^\mathrm{s}$ & $033^\circ 22'$ \\
Skymapper & $00^\mathrm{h} 15^\mathrm{m} 43^\mathrm{s}$ & $033^\circ 08'$ \\
TAROT & $00^\mathrm{h} 25^\mathrm{m} 06^\mathrm{s}$ & $027^\circ 16'$ \\
TAROT-S & $00^\mathrm{h} 04^\mathrm{m} 27^\mathrm{s}$ & $030^\circ 33'$ \\
ROTSE III-a & $00^\mathrm{h} 34^\mathrm{m} 29^\mathrm{s}$ & $028^\circ 54'$ \\
ROTSE III-b & $21^\mathrm{h} 06^\mathrm{m} 13^\mathrm{s}$ & $-45^\circ 30'$ \\
ROTSE III-c & $00^\mathrm{h} 43^\mathrm{m} 51^\mathrm{s}$ & $022^\circ 20'$ \\
ROTSE III-d & $00^\mathrm{h} 16^\mathrm{m} 39^\mathrm{s}$ & $027^\circ 16'$ \\
Zadko & $00^\mathrm{h} 25^\mathrm{m} 06^\mathrm{s}$ & $032^\circ 39'$ \\
\tableline
\end{tabular}
 \\~\\~\\
				(c) Table of pointings
			\end{center}
		\end{minipage}
		\includegraphics{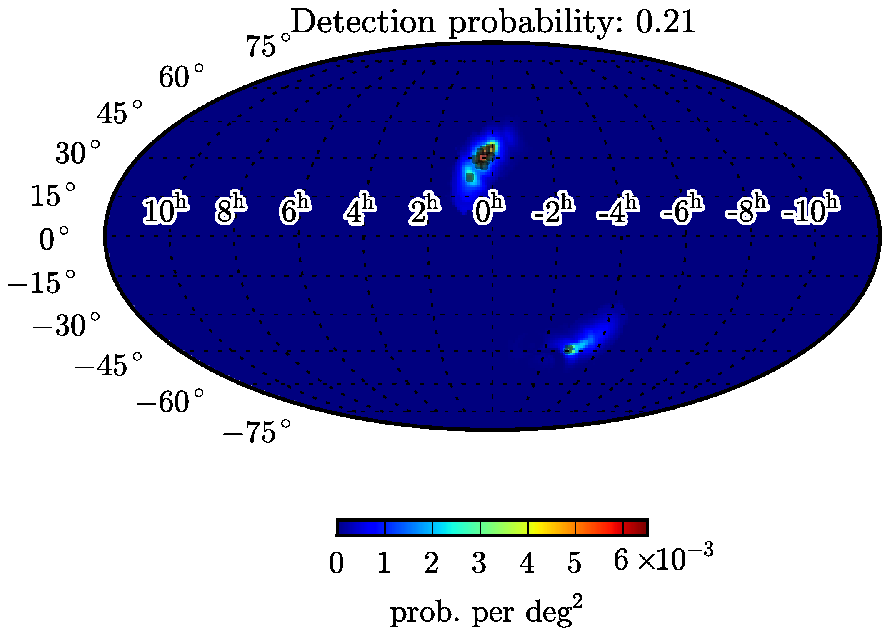} \\
		(d) All sky
	\end{center}
	\caption{\label{fig:demonstration}Demonstration of a \ac{BAYESTAR} observation plan for a simulated \ac{NS}\nobreakdashes--\ac{NS} \ac{CBC} event injected into synthetic initial \acs{LIGO} noise.  The event produced triggers in the H1, L1, and V1 detectors.  The \ac{GW} sky map, $p(\boldsymbol\omega | \textsc{gw})$, in posterior probability per square degree, is shown in color in the online version and in shades of gray in the print version.  The \acp{FOV} of the telescopes are shown as shaded gray rectangles.  Panels (a) and (b) show Lambert azimuthal equal area projections of two $30^\circ \times 30^\circ$ patches where observations were planned.  Panel (c) provides the selected telescope pointings in tabular form.  Panel (d) shows the selected observations on an all\nobreakdashes-sky Mollweide projection.}
\end{figure*}

We compared the imaging efficiency of our three multiple telescope planners
using a set of 2126 \ac{GW} sky maps from signals injected into simulated
initial \acs{LIGO} and Virgo noise over a period of 24 hours.  The simulated \ac{GW} signals are accurate to second post\nobreakdashes-Newtonian order in phase.  Component masses are drawn from a uniform distribution from 1 to 15~$M_\odot$ with total mass is restricted to $<$20~$M_\odot$.  Distances are drawn randomly from a log\nobreakdashes-uniform distribution from 10 to 40~Mpc.
A sky map was produced for each injection using the time delay method described in \citet{CBCLowLatency}.  We generated two observing plans for each sky map for each of the three planners.  The first observing plan used $v_i(\boldsymbol\omega) = 1$ for all observations $i$ and sky locations $\boldsymbol\omega$, disregarding the positions of the sun and Earth.  The second observing plan used $v_i(\boldsymbol\omega) = 1$ above the horizon and at all sky locations from which the sun was at least 18$^\circ$ below the horizon, and zero elsewhere, simulating observing conditions on an atmosphereless, spherical Earth.  The output of the \emph{greedy\nobreakdashes-sorted} planner was used as the initial proposal for simulated annealing.  Pi of the Sky and Liverpool Telescope were excluded from the simulation because the areas of their \acp{FOV} are outliers with regard to the other \acp{FOV} of $\sim$1\nobreakdashes--10~deg$^2$ represented in the model.  Figure~\ref{fig:demonstration} illustrates an observing plan generated with the greedy algorithm with sun and horizon constraints disabled.

\begin{figure}
	\plotone{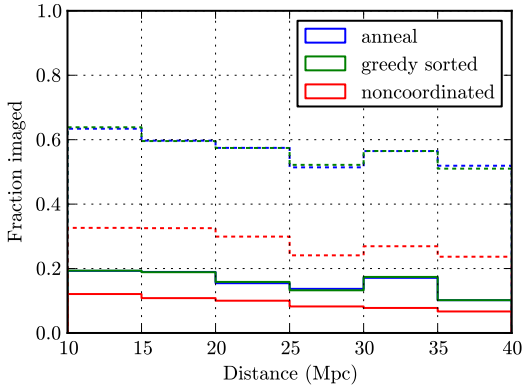}
	\caption{\label{fig:efficiency_distance}Average detection efficiency of \emph{noncoordinated}, \emph{greedy\nobreakdashes-sorted}, and \emph{anneal} planners, for sources located within spherical shells 5~Mpc thick at 10\nobreakdashes--40~Mpc.  Solid lines include sun and horizon interference; dashed lines ignore interference considerations.}
\end{figure}

Figure~\ref{fig:efficiency_distance} shows the measured imaging efficiency as a function of the luminosity distance.  For each sky map and for each planner, we tested whether the injected location would have been imaged, i.e., whether the injected source location was inside any of the \acp{FOV} of the telescopes.  Dividing the injection set into six concentric shells, each 5~Mpc deep, from 10 to 40~Mpc, we tallied the number of sources that would have been imaged and the total number of sources.  Because the injections were drawn from a uniform distribution in $\log D_L$, we weighted each injection by
\begin{equation}
	{D_L}^2 \left| \frac{\partial \log D_L}{\partial D_L} \right|^{-1} \equiv {D_L}^3
\end{equation}
in order to approximate a uniform\nobreakdashes-in\nobreakdashes-volume distribution.  The dashed lines represent the observing plans that disregard the presence of the sun and the Earth whereas the solid lines represent the observing plans that fully account for these effects.

\begin{figure}
	\plotone{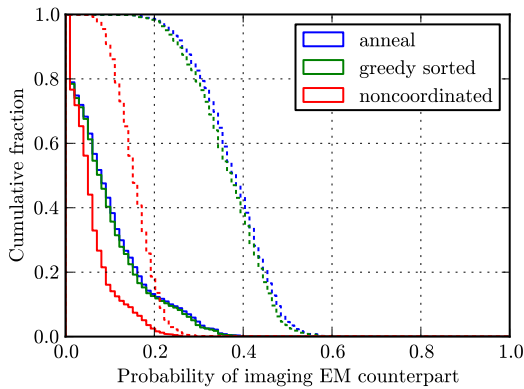}
	\caption{\label{fig:efficiency_histogram}Histogram of expected detection efficiency $p_\mathrm{any}$ as computed using Equation~(\ref{eq:p_any}), for sources uniformly distributed in volume from 10 to 40~Mpc.  As in Figure~\ref{fig:efficiency_distance}, solid lines include sun and horizon interference and dotted lines ignore interference considerations.}
\end{figure}

For each injection, we also computed the probability $p_\mathrm{any}$ of imaging the source, marginalized over $(\theta, \phi)$ and not making use of knowledge of the true sky location, using Equation~(\ref{eq:p_any}).  We then formed a histogram of $p_\mathrm{any}$ over all of our injections, again weighting by the correction factor $D_L^3$.  This histogram is shown in Figure~\ref{fig:efficiency_histogram}.

Two systematic effects are apparent in Figure~\ref{fig:efficiency_distance}.  First, for all planning algorithms, accounting for the sun results in a reduction in efficiency by a factor of $\approx 1/3$, independent of distance.  This is because astronomical twilight begins and ends when the sun is 18$^\circ$ below the horizon \citeyearpar[Astronomical Almanac][]{u2010astronomical}; solar illumination is faint enough for astronomical observations from the Earth's surface over over a fraction $\left(1 - \cos \left[90^\circ-18^\circ\right] \right) / 2 \approx 0.35$ of the sky.  Second, the efficiency falls as luminosity distance increases because the area of the \ac{GW} error region grows as \acl{SNR} decreases.

Both Figures~\ref{fig:efficiency_distance}~and~\ref{fig:efficiency_histogram} demonstrate that coordination of many optical telescopes increases the chances of imaging the true location of the \ac{GW} source.  Whether or not sun and horizon interference are included, the \emph{anneal} and \emph{greedy\nobreakdashes-sorted} planners achieve about twice the detection efficiency of the \emph{noncoordinated} planner for the network of telescopes used in the last joint \acs{LIGO}\nobreakdashes--Virgo science run and in our model.

From Figure~\ref{fig:efficiency_histogram}, we see that simulated annealing generally outperforms the greedy algorithm.  At least a slight improvement is expected because the observing plan generated by the \emph{greedy\nobreakdashes-sorted} algorithm is used as the initial proposal for simulated annealing.  Simulated annealing would also give only a slight improvement if the \emph{greedy\nobreakdashes-sorted} observing plan was already nearly optimal.

\section{Discussion}

We have presented a framework for planning optimal \ac{EM} follow\nobreakdashes-up of \ac{GW} transients.  We have shown that, if each telescope images one sky location, a coordinated approach with a particular
network of telescopes is twice as likely to image the true location
of a \ac{GW} event, versus an uncoordinated approach.  It remains to be seen whether coordination will be important for more realistic observing strategies, which will involve taking multiple images with each telescope.  A successful observational program also will likely involve several different stages calling for markedly different kinds of instruments.  In the context of a grand strategy for the follow\nobreakdashes-up of \ac{GW} candidates \citep{MetzgerBerger:2012}, spatially coordinated observing campaigns like those we have considered are most likely to be applicable to prompt, rapidly fading, on\nobreakdashes-axis afterglows, though it is unclear whether coordination will be relevant for later-rising and fainter counterparts.  This will be the subject of future work.

We have found that a simple greedy algorithm performs almost as well as
simulated annealing while taking much less time on average.  In the interest of
low latency, we have provided a parallel implementation of the convolution
algorithm so that observation plans can be generated rapidly on a multicore
workstation.  The observation planning code has been operated in a low latency
mode during software engineering runs of the joint \acs{LIGO}\nobreakdashes--Virgo low latency
data analysis.

In this work, the objective function that we have used is the probability of
\emph{imaging} the source.  This is a proxy for the more desirable objective
function of the probability of \emph{detecting} the source, which must take into
account the limiting magnitudes of the observations, a description of the potential false positives, and a model of the \ac{EM} emission.  A weighting
could be assigned to each telescope's \ac{FOV} based upon the integrated flux
that the telescope would be able to gather given the anticipated power law index
of the light curve and the telescope's limiting magnitude, slew time, and
exposure time.  For example, assuming a power\nobreakdashes-law light curve $F_\nu \propto t^{-\alpha}$, at time $t$ a telescope with limiting magnitude $M$ would be able to
detect a source up to a maximum luminosity distance $D_{L,\mathrm{max}}
\propto \sqrt{2.512^M t^{-\alpha}}$.  We can marginalize over the unknown distance to
the source, assuming that nearby \acp{GRB} are uniformly distributed in space.
Neglecting cosmological corrections, the detection probability of a single
telescope might be expressed by
\begin{equation}
	p(\textsc{em}_i|\boldsymbol\gamma_i, \boldsymbol\omega) \propto \begin{cases}
		\left(2.512^{M_i} {t_i}^{-\alpha}\right)^{3/2} & \text{ inside \ac{FOV}}, \\
		0 & \text{ outside \ac{FOV}},
	\end{cases}
\end{equation}
where $t_i$ is the time of observation $i$ and $M_i$ is the limiting magnitude of observation $i$.  We could also marginalize over the power\nobreakdashes-law index $\alpha$ and the initial luminosity.

Although this prescription will be appropriate for a single telescope, for
multiple telescopes or multiple pointings of the same telescope it will be more
appropriate to form the probability of detection with the network of telescopes,
conditioned on the luminosity distance, and \emph{then} marginalize over luminosity
distance.  Cast in this form, the multiple telescope detection probability is
\begin{multline}
	p_\mathrm{any} = 1 - \frac{1}{V} \int \prod_i \bigg[ 1 - \Theta\big[D_{L,\mathrm{max},i} \left(L_i, M_i\right) - D_L\big]
		b_i \left( R^{-1}[\boldsymbol\gamma_i]\, \boldsymbol\omega \right) v_i(\boldsymbol\omega) \bigg]
		s(\boldsymbol\omega) \\
		p(L_1, L_2, \dots, L_N) \, \mathrm{d}L_1 \mathrm{d}L_2 \dots \mathrm{d}L_N \;\,
		{D_L}^2 \mathrm{d}D_L \, \mathrm{d}\boldsymbol{\Omega}
\end{multline}
where $V$ is the volume of integration and $\Theta$ is the Heaviside step function.  $D_{L,\mathrm{max},i} \propto \sqrt{2.512^{M_i} L_i}$ is the farthest detectable luminosity distance for observation $i$, being a function of the source's luminosity $L_i$ at the time $t_i$ of the observation and the limiting magnitude $M_i$.  This figure of merit is not as simple as
Equation~(\ref{eq:p_any}), but it may behave better under simulated annealing
because the \acp{FOV} no longer have to move rigidly around each other without
overlapping.  This is also the subject of future work.

\acknowledgements Source code for \ac{BAYESTAR} is available on the \acs{LIGO} \acl{DASWG} web site at \url{http://www.lsc-group.phys.uwm.edu/daswg/projects/bayestar.html}.

Some of the results in this paper have been derived using HEALPix \citep{healpix}.

\acs{LIGO} was constructed by the California Institute of Technology and Massachusetts Institute of Technology with funding from the \ac{NSF} and operates under cooperative agreement PHY\nobreakdashes-0107417.  Some results were produced on the NEMO computing cluster operated by the Center for Gravitation and Cosmology at University of Wisconsin\nobreakdashes--Milwaukee under \ac{NSF} Grants PHY\nobreakdashes-0923409 and PHY\nobreakdashes-0600953.  This research is supported by the \ac{NSF} through a Graduate Research Fellowship to L.S.  A.S. acknowledges support from the \acs{LIGO} Laboratory Summer
Undergraduate Research Fellowship, Visitors, and \ac{NSF} \ac{REU} programs.  This paper has \acs{LIGO} Document Number \acs{LIGO}\nobreakdashes-P1200033\nobreakdashes-v4.

\appendix
\section{Implementation}

Equation~(\ref{eq:optimal-pointing-single-telescope}) is reminiscent of a convolution integral.  This leads us to think of $b_i(\boldsymbol\omega)$ as an analogue of a \ac{FIR} filter, but on the unit sphere.  A \ac{FIR} filter with a sufficiently compact kernel can be efficiently evaluated in so\nobreakdashes-called direct form by advancing the kernel one sample point at a time and taking an inner product with the signal.  On the other hand, a \ac{FIR} filter with a long, extended kernel is generally more efficiently implemented using frequency domain techniques such as the classic overlap\nobreakdashes-save algorithm~\citep[Numerical Recipes,][]{numerical-recipes-chapter-13}, which exploits the convolution theorem and the \ac{FFT}.

Similarly, if the telescope's \ac{FOV} spans sufficiently little area, then a discretized form of Equation~(\ref{eq:optimal-pointing-single-telescope}) can be efficiently evaluated by subdividing both the \ac{FOV} and the \ac{GW} sky map into pixels, and for each pointing $\boldsymbol\gamma_i$, rotating the pixels in the kernel and then taking their inner product with the pixels in the \ac{GW} sky map.  Transforming just the nonzero pixels of the kernel results in a substantial computational speedup; for a $\sim$1~deg$^2$ \ac{FOV}, this reduces the number of calculations by a factor $4(180)^2 / \pi \sim 4 \times 10^{4}$.

If, however, the telescope's \ac{FOV} subtends a large solid angle, then the ``direct'' algorithm outlined above will be computationally expensive, with a cost growing as $\mathcal{O}(\npix^2)$, where $\npix$ is the number of pixels covering the entire unit sphere.  The \ac{FFT}\nobreakdashes-based frequency domain techniques for \ac{FIR} filters also have analogous fast \emph{multipole} transform algorithms for functions defined on $S^2$.  Considerable study has been devoted to fast multipole methods because of their utility in solving differential equations with approximate spherical symmetry.  Example applications can be found in fields as varied as \ac{CMB} mapping \citep{healpix}, numerical astrophysics, quantum chemistry, and even the entertainment industry~\citep{mcewen}.  Presently, the best known upper bound on the complexity of both spherical harmonic transforms and fast convolution on the sphere \citep{healpix,wandelt} is $\mathcal{O}(\npix^{3/2}) \sim \mathcal{O}(L^3)$, where $L \sim \mathcal{O}(\npix^{1/2})$ is the azimuthal index of the highest non\nobreakdashes-vanishing (or non\nobreakdashes-negligible) spherical harmonic component.

For a sufficiently dense pixel resolution and sufficiently large \ac{FOV}, the multipole method will inevitably be more efficient than the direct method.  Anticipating that one approach or the other might be more computationally efficient for a given \ac{FOV}, we implemented both a ``direct'' algorithm and a ``multipole'' algorithm, described below.

\subsection{``Direct'' algorithm}

For the ``direct'' algorithm, it is important that the sky is divided into patches, or pixels, of equal area.  The pixelization scheme defines a surjective map from points on the unit sphere to integer pixel indices, $f\colon S^2 \mapsto \mathbb{Z}_\npix$.  It is desirable to be able to compute the discretized version of the rotated \ac{FOV}, $b_i(\boldsymbol\gamma_i^{-1} \boldsymbol\omega)$, by transforming the pixel coordinates themselves.  If some pixels are much larger or smaller than others, then some rotations will require a great deal of sub\nobreakdashes-pixel interpolation.  An example of a particularly troublesome pixelization scheme would be a division into a equiangular lattice in right ascension and declination, because pixels near the poles would subtend much smaller solid angles than pixels near the equator.  It is much better if the pixels are of equal or nearly equal size, for then a ``nearest neighbor'' interpolation scheme consists of nothing more than a permutation of the pixel indices.  Such an interpolation scheme ought to converge to first order in the pixel area, while being exceptionally fast.

\begin{algorithm}
\caption{``Direct'' convolution for single telescope case.}
\begin{algorithmic}
	\STATE $\{s_i[0], s_i[1], \dots, s_i[\npix-1]\} \gets \textit{pixelized map}$
	\STATE $\{v_i[0], v_i[1], \dots, v_i[\npix-1]\} \gets \textit{pixelized mask}$
	\STATE $\{b_i[0], b_i[1], \dots, b_i[\npix-1]\} \gets \textit{pixelized \ac{FOV}}$
	\STATE $\{p_i[0], p_i[1], \dots, p_i[\npix-1]\} \gets \{0, 0, \dots, 0\}$
	\FOR{$j = 0 \to \npix - 1$}
		\STATE $\boldsymbol\gamma \gets \mathrm{pix2ang}\, (j)$
		\FOR{$k \in \mathbb{Z} \colon 0 \leq k < \npix, b_i[k] \neq 0$}
			\STATE $\boldsymbol\omega \gets \mathrm{pix2ang}\, (k)$
			\STATE $k' \gets \mathrm{ang2pix}\, (R[\boldsymbol\gamma^{-1}] \boldsymbol\omega)$
			\STATE $p_i[j] \gets p_i[j] + b_i[k] \times v_i[k'] \times s_i[k']$
		\ENDFOR
	\ENDFOR
\end{algorithmic}
\end{algorithm}

For this reason, we chose to build an implementation of the ``direct'' algorithm in HEALPix~\citep{healpix}, a special pixelization of the sphere and an accompanying software package that has become one of the workhorses of all\nobreakdashes-sky astronomical data analysis.  HEALPix stands for Hierarchical Equal Area isoLatitude Pixelization.  As the name says, it has three chief virtues.  The first is that successive levels of resolution are derived by subdividing pixels.  This makes it straightforward to downsample (decreasing resolution) or upsample (increase resolution) a map.  Also, for search operations it provides a log\nobreakdashes-linear speedup that is analogous to binary searches trees or octrees.  The second virtue is that all HEALPix pixels at a given resolution have the same area, satisfying the requirement for our ``direct'' convolution algorithm.  Finally, the isolatitude property, that pixels are arranged on rings of constant latitude, facilitates computationally efficient multipole transforms at any resolution, which we exploit in the next section.

\subsection{``Multipole'' algorithm}

The ``multipole'' algorithm is the analog of \ac{FFT} convolution in spherical coordinates.  Our implementation of fast convolution with HEALPix is based on \cite{wandelt}.  A weighted outer product is formed from the coefficients of the spherical harmonic expansions of the sky map and the telescope \ac{FOV}.  The resultant tensor is the 2D \ac{DFT} of the convolution on a equiangular grid in $(\theta, \phi)$.  Our contribution is to specify a computationally efficient procedure for obtaining the convolution on the HEALPix grid instead of the equiangular grid in spherical coordinates.  This is an especially useful feature for our application because output in HEALPix coordinates facilitates further spatial manipulation after the convolution.  A detailed explanation of the algorithm is provided below.  Source code in C++ is provided in the supplementary material.

Without loss of generality, from here on we drop the environmental factor $v_i (\boldsymbol\omega)$ and the telescope index $i$, leaving just
$$
	p(\textsc{em} | \boldsymbol\gamma, \textsc{gw}) = \int s(\boldsymbol\omega) b\left(R^{-1}[\boldsymbol\gamma]\, \boldsymbol\omega\right) \, \mathrm{d}\boldsymbol{\Omega}.
$$
Though the integral involves a \emph{passive} rotation of the beam by $R^{-1}[\boldsymbol\gamma]$, it is conceptually simpler to think of this is as an \emph{active} rotation by $R[\boldsymbol\gamma]$.  In the language of spherical harmonics, this integral can be evaluated at any $\boldsymbol\gamma = (\theta, \phi)$ through the sum
\begin{align}
	\label{eq:single-telescope-multipole-sum}
	p(\textsc{em} | \theta, \phi, \textsc{gw}) = \sum_{l=0}^L \sum_{m=-l}^l \sum_{n=-l}^l s_{lm}^* D^l_{mn} (\phi, \theta, 0) b_{ln}
\end{align}
where
\begin{equation}
	D^l_{mn} (\alpha, \beta, \gamma) = e^{-i m \alpha} d^l_{mn} (\beta) e^{-i n \gamma}
\end{equation}
is the Wigner matrix element denoting the active rotation first by $\gamma$ about the $z$\nobreakdashes-axis, then by $\beta$ about the new $y$\nobreakdashes-axis, and finally about $\alpha$ about the final $z$\nobreakdashes-axis.  (\citet{sakurai1994modern} use the active convention; \citet{edmonds1996angular} and \emph{Mathematica} employ the passive convention.)  $s_{lm}$ and $b_{ln}$ are the spherical harmonic coefficients of $s(\boldsymbol\omega)$ and $b(\boldsymbol\omega)$, respectively, assumed to be zero or at least negligibly small for $l > L$.  $d^l_{mn} (\beta)$ is the reduced Wigner matrix element representing an active rotation about the $y$\nobreakdashes-axis by $\beta$.  The basic idea is to employ a Fourier expansion of the reduced Wigner coefficients \citep{edmonds1996angular, risbo, wandelt},
\begin{equation}
	d^l_{mn} (\beta) = i^{m-n} \sum_{m' = -l}^l \Delta^l_{m'm} \Delta^l_{m'n} e^{-i m \beta},
\end{equation}
where $\Delta^l_{mn} = d^l_{mn} (\pi/2)$, to convert the summation in Equation~(\ref{eq:single-telescope-multipole-sum}) to a two\nobreakdashes-dimensional, discrete Fourier transform,
\begin{align}
	p(\textsc{em} | \theta, \phi, \textsc{gw}) &= \sum_{l=0}^L \sum_{m=-l}^l \sum_{m'=-l}^l \sum_{n=-l}^l i^{m-n} s_{lm}^* \Delta^l_{m'm} \Delta^l_{m'n} e^{-i m' \theta} e^{-i m \phi} b_{ln} \nonumber\\
		&\equiv \sum_{m=-L}^L \sum_{m'=-L}^L T_{m m'} e^{-i m' \theta} e^{-i m \phi}
\end{align}

First, compute both $s_{lm}$ and $b_{lm}$ using the HEALPix routine \texttt{map2alm} for $m \geq 0$.  Since $s(\boldsymbol\omega)$ and $b(\boldsymbol\omega)$ are purely real, the remaining coefficients $m < 0$ are determined by conjugate symmetry, $s_{l,-m} = (-1)^m s_{lm}^*$.  Also, all of the $m=0$ coefficients are purely real.

Next, compute $\bar{s}_{lm} = (-i)^m s_{lm}$ and $\bar{b}_{lm} = (-i)^m b_{lm}$ for $m \geq 0$ to transfer the phase factor $i^{m-n}$ to the beam and sky map.

Next, evaluate the sum over $n$ in $\mathcal{O}(L^3)$ time, yielding $X_{lm'}$:
\begin{equation}
	X_{lm'} \equiv \sum_{n=-l}^l \Delta^l_{m'n} \bar{b}_{ln} = \Delta^l_{m'0} b_{l0} + 2 \sum_{m=1}^l \Delta^l_{m'n}
		\begin{cases}
			\re \left[\bar{b}_{ln}\right] & \text{if } (l - m - n) \text{ is even} \\
			\im \left[\bar{b}_{ln}\right] & \text{if } (l - m - n) \text{ is odd}
		\end{cases}
\end{equation}
The expression after the equality sign makes use of $\bar{b}_{ln} = \bar{b}_{l,-n}^*$ and the identity \citep{mcewen} $d^l_{mn}(\pi - \beta) = (-1)^{l-m} d^l_{m,-n}(\beta) \Longrightarrow \Delta^l_{mn} = (-1)^{l-m} \Delta^l_{m,-n}$.

Next, for $0 \leq m \leq L$ and $-L \leq m' \leq L$, the sum over $l$ is computed, also in $\mathcal{O}(L^3)$ time:
\begin{equation}
	T_{mm'} \equiv \sum_{\mathclap{l=\max (|m|, |m'|)}}^L \bar{s}_{lm}^* \Delta^l_{m'm} X_{lm'} \\
		= \sum_{\mathclap{l=\max (m, |m'|)}} \bar{s}_{lm}^* \Delta^l_{|m'|m} \begin{cases}
			X_{lm'} & \text{if } m' \geq 0 \\
			X_{l|m'|}^* (-1)^{l-m-m'} & \text{if } m' < 0
		\end{cases}.
\end{equation}

At this point, it is possible to recover the convolution on a equiangular grid on right ascension and declination by performing a 2D forward Fourier transform of $T_{mm'}$.  However, this grid will oversample the poles; it is preferable to obtain the convolution at the HEALPix pixel sites.  The HEALPix grid is arranged such that the pixel centers $(\theta_j, \phi_{jk})$ lie on a set of $\mathcal{O}(L)$ iso(co)latitude rings at colatitudes $\theta_j$.  For the colatitude $\theta_j$ particular to each ring, the sum over $m'$ is performed in a \emph{total} of $\mathcal{O}(L^3)$ operations, yielding
\begin{equation}
	T_m(\theta_j) \equiv \sum_{m'=-L}^L T_{mm'} e^{-i m' \theta_j}.
\end{equation}
This 1D Fourier series, $T_m(\theta_j)$, is now shifted in phase so that $T_0(\theta_j)$ is the value at the first pixel in the ring centered at $(\phi_{j0}, \theta_j)$, resulting in
\begin{equation}
	T'_m(\theta_j) \equiv T_m(\theta_j) e^{-i m \phi_{j0}}.
\end{equation}
Since the convolution must be everywhere real, we know that $T'_{-m}(\theta_j) = [T'_m(\theta_j)]^*$.

Now, in rings close to the poles, the number of pixels per ring $n_{\mathrm{ring},j}$ is small; we assume that in \emph{all} rings $n_{\mathrm{ring},j} < 2 L + 1$.  For more efficient evaluation, the Fourier series
can be folded, or aliased, to $T''_m(\theta_j)$, for $0 \leq m < n_{\mathrm{ring},j} / 2$, such that
\begin{equation}
	\label{eq:ring-dft}
	p(\textsc{em} | \theta_j, \phi_{jk}, \textsc{gw}) = \sum_{m = -L}^L T'_m (\theta_j) \exp \left[ -\frac{2 \pi i k m}{n_{\mathrm{ring},j}} \right]
	= \sum_{m=-(n_{\mathrm{ring},j}/2-1)}^{n_{\mathrm{ring},j}/2-1} T''_m(\theta_j) \exp \left[ -\frac{2 \pi i k m}{n_{\mathrm{ring},j}} \right]
\end{equation}
where
\begin{equation}
	T''_m(\theta_j) = \sum_i \left( T'_{m + n_{{\mathrm{ring},j}} i} + T'_{m - n_{\mathrm{ring},j} i} \right) \text{ for } 0 \leq m < n_{\mathrm{ring},j} / 2.
\end{equation}
The Fourier series can be summed directly in a total of $\mathcal{O} (L^3)$ operations for all rings.  Although not necessary to preserve the overall cubic scaling with $L$, this final step can be sped up by employing a \ac{FFT} instead of a direct sum.  \ac{FFT} packages such as fftw\footnote{\url{http://www.fftw.org/}} conventionally provide a \emph{real\nobreakdashes-to\nobreakdashes-complex} forward \ac{FFT} for real\nobreakdashes-valued input data, producing just the nonnegative frequency components, and a corresponding inverse \emph{complex\nobreakdashes-to\nobreakdashes-real} \ac{FFT} converting just nonnegative frequency components back to real data.  We would like to use a \emph{complex\nobreakdashes-to\nobreakdashes-real} transform to fill in the ring pixels from the Fourier series, but the negative sign in the exponent of Equation~(\ref{eq:ring-dft}) makes it a forward, rather than an inverse, transform.  Application of a \emph{complex\nobreakdashes-to\nobreakdashes-real} \ac{FFT} would result in the ring pixels coming out in reversed order.  For this reason, we compute
\begin{equation}
	T'''_m(\theta_j) = \left[T''_m(\theta_j)\right]^*
\end{equation}
and then perform the \emph{complex\nobreakdashes-to\nobreakdashes-real} \ac{FFT} on \emph{this} to fill in the pixels in each ring.

\subsection{Comparison}

We implemented both the ``direct'' and ``multipole'' convolution algorithms in C/C++.  The outermost loops are accelerated using OpenMP\footnote{\url{http://openmp.org/}}, a widely supported compiler extension for generating parallel code for multiprocessor machines.  Much of the HEALPix library itself, including the spherical harmonic transform routine \texttt{map2alm}, is also accelerated with OpenMP.  For the multipole algorithm, the final \ac{DFT}s are performed using fftw.

Operating with one thread on a quad core Sun Fire X2200 M2 and a pixel resolution of $\approx$0.05~deg$^2$, the direct algorithm ran in 8.9~s for the \acs{ROTSE}~III \ac{FOV}, 38.4~s for \acs{QUEST}, and 931.2~s for Pi of the Sky South.  The multipole algorithm, whose run time does not depend on the \ac{FOV}, took 23.4~s.  Using 8 threads, \acs{ROTSE}\nobreakdashes-III took 1.3~s, \acs{QUEST} 5.4~s, Pi of the Sky South 128.9~s, and the multipole algorithm took 6.6~s.

\bibliographystyle{apj}
\bibliography{proposal,telescope,telescope_list}

\begin{thebibliography}{71}
\expandafter\ifx\csname natexlab\endcsname\relax\def\natexlab#1{#1}\fi

\bibitem[{Aasi {et~al.}(2012{\natexlab{a}})Aasi, Abadie, Abadie, \&
  et~al.}]{OpticalImageAnalysis}
Aasi, J., Abadie, J., Abadie, B.~P., \& et~al. 2012{\natexlab{a}}, in
  preparation

\bibitem[{Aasi {et~al.}(2012{\natexlab{b}})Aasi, Abadie, Abadie, \&
  et~al.}]{SwiftFollowup}
---. 2012{\natexlab{b}}, LIGO-P1100038-v4, in preparation

\bibitem[{Abadie {et~al.}(2010)}]{rates}
Abadie, J., {et~al.} 2010, Class. Quantum Grav., 27, 173001

\bibitem[{Abadie {et~al.}(2011)Abadie, Abbott, Abbott, Abernathy, Accadia,
  Acernese, Adams, Adhikari, Affeldt, Allen, Allen, Ceron, Amariutei, Amin,
  Anderson, Anderson, Antonucci, Arai, Arain, Araya, Aston, Astone, Atkinson,
  Aufmuth, Aulbert, Aylott, Babak, Baker, Ballardin, Ballmer, Barker, Barnum,
  Barone, Barr, Barriga, Barsotti, Barsuglia, Barton, Bartos, Bassiri,
  Bastarrika, Basti, Bauchrowitz, Bauer, Behnke, Beker, Bell, Belletoile,
  Belopolski, Benacquista, Bertolini, Betzwieser, Beveridge, Beyersdorf,
  Bilenko, Billingsley, Birch, Birindelli, Biswas, Bitossi, Bizouard, Black,
  Blackburn, Blackburn, Blair, Bland, Blom, Bock, Bodiya, Bogan, Bondarescu,
  Bondu, Bonelli, Bonnand, Bork, Born, Boschi, Bose, Bosi, Bouhou, Boyle,
  Braccini, Bradaschia, Brady, Braginsky, Brau, Breyer, Bridges, Brillet,
  Brinkmann, Brisson, Britzger, Brooks, Brown, Brummit, Budzyński, Bulik,
  Bulten, Buonanno, Burguet-Castell, Burmeister, Buskulic, Buy, Byer, Cadonati,
  Cagnoli, Cain, Calloni, Camp, Campagna, Campsie, Cannizzo, Cannon, Canuel,
  Cao, Capano, Carbognani, Caride, Caudill, Cavaglià, Cavalier, Cavalieri,
  Cella, Cepeda, Cesarini, Chaibi, Chalermsongsak, Chalkley, Charlton,
  Chassande-Mottin, Chelkowski, Chen, Chincarini, Christensen, Chua, Chung,
  Chung, Clara, Clark, Clark, Clayton, Cleva, Coccia, Colacino, Colas, Colla,
  Colombini, Conte, Cook, Corbitt, Cornish, Corsi, Costa, Coughlin, Coulon,
  Coward, Coyne, Creighton, Creighton, Cruise, Culter, Cumming, Cunningham,
  Cuoco, Dahl, Danilishin, Dannenberg, D'Antonio, Danzmann, Das, Dattilo,
  Daudert, Daveloza, Davier, Davies, Daw, Day, Dayanga, Rosa, DeBra,
  Debreczeni, Degallaix, del Prete, Dent, Dergachev, DeRosa, DeSalvo,
  Dhurandhar, Fiore, Lieto, Palma, Emilio, Virgilio, Díaz, Dietz, Donovan,
  Dooley, Dorsher, Douglas, Drago, Drever, Driggers, Dumas, Dwyer, Eberle,
  Edgar, Edwards, Effler, Ehrens, Engel, Etzel, Evans, Evans, Factourovich,
  Fafone, Fairhurst, Fan, Farr, Fazi, Fehrmann, Feldbaum, Ferrante, Fidecaro,
  Finn, Fiori, Flaminio, Flanigan, Foley, Forsi, Forte, Fotopoulos, Fournier,
  Franc, Frasca, Frasconi, Frede, Frei, Frei, Freise, Frey, Fricke, Friedrich,
  Fritschel, Frolov, Fulda, Fyffe, Galimberti, Gammaitoni, Garcia, Garofoli,
  Garufi, Gáspár, Gemme, Genin, Gennai, Ghosh, Giaime, Giampanis, Giardina,
  Giazotto, Gill, Goetz, Goggin, González, Gorodetsky, Goßler, Gouaty, Graef,
  Granata, Grant, Gras, Gray, Greenhalgh, Gretarsson, Greverie, Grosso, Grote,
  Grunewald, Guidi, Guido, Gupta, Gustafson, Gustafson, Hage, Hallam, Hammer,
  Hammond, Hanks, Hanna, Hanson, Harms, Harry, Harry, Harstad, Hartman,
  Haughian, Hayama, Hayau, Hayler, Heefner, Heitmann, Hello, Hendry, Heng,
  Heptonstall, Herrera, Hewitson, Hild, Hoak, Hodge, Holt, Hong, Hooper,
  Hosken, Hough, Howell, Huet, Hughey, Husa, Huttner, Ingram, Inta, Isogai,
  Ivanov, Jaranowski, Johnson, Jones, Jones, Jones, Ju, Kalmus, Kalogera,
  Kandhasamy, Kanner, Katsavounidis, Katzman, Kawabe, Kawamura, Kawazoe, Kells,
  Kelner, Keppel, Khalaidovski, Khalili, Khazanov, Kim, Kim, King, Kinzel,
  Kissel, Klimenko, Kondrashov, Kopparapu, Koranda, Korth, Kowalska, Kozak,
  Kringel, Krishnamurthy, Krishnan, Królak, Kuehn, Kumar, Kwee, Landry, Lantz,
  Lastzka, Lazzarini, Leaci, Leong, Leonor, Leroy, Letendre, Li, Li, Liguori,
  Lindquist, Lockerbie, Lodhia, Lorenzini, Loriette, Lormand, Losurdo, Lu,
  Luan, Lubinski, Lück, Lundgren, Macdonald, Machenschalk, MacInnis,
  Mageswaran, Mailand, Majorana, Maksimovic, Man, Mandel, Mandic, Mantovani,
  Marandi, Marchesoni, Marion, M\'arka, M\'arka, Maros, Marque, Martelli,
  Martin, Martin, Marx, Mason, Masserot, Matichard, Matone, Matzner, Mavalvala,
  McCarthy, McClelland, McGuire, McIntyre, McKechan, Meadors, Mehmet, Meier,
  Melatos, Melissinos, Mendell, Mercer, Merill, Meshkov, Messenger, Meyer,
  Miao, Michel, Milano, Miller, Minenkov, Mino, Mitrofanov, Mitselmakher,
  Mittleman, Miyakawa, Moe, Moesta, Mohan, Mohanty, Mohapatra, Moraru, Moreno,
  Morgado, Morgia, Mosca, Moscatelli, Mossavi, Mours, Mow-Lowry, Mueller,
  Mukherjee, Mullavey, Müller-Ebhardt, Munch, Murray, Nash, Nawrodt, Nelson,
  Neri, Newton, Nishida, Nishizawa, Nocera, Nolting, Ochsner, O'Dell, Ogin,
  Oldenburg, O'Reilly, O'Shaughnessy, Osthelder, Ott, Ottaway, Ottens,
  Overmier, Owen, Page, Pagliaroli, Palladino, Palomba, Pan, Pankow, Paoletti,
  Papa, Parameswaran, Pardi, Parisi, Pasqualetti, Passaquieti, Passuello,
  Patel, Pathak, Pedraza, Pekowsky, Penn, Peralta, Perreca, Persichetti,
  Phelps, Pichot, Pickenpack, Piergiovanni, Pietka, Pinard, Pinto, Pitkin,
  Pletsch, Plissi, Podkaminer, Poggiani, Pöld, Postiglione, Prato, Predoi,
  Price, Prijatelj, Principe, Privitera, Prix, Prodi, Prokhorov, Puncken,
  Punturo, Puppo, Quetschke, Raab, Rabeling, Rácz, Radkins, Raffai, Rakhmanov,
  Ramet, Rankins, Rapagnani, Raymond, Re, Redwine, Reed, Reed, Regimbau, Reid,
  Reitze, Ricci, Riesen, Riles, Roberts, Robertson, Robinet, Robinson,
  Robinson, Rocchi, Roddy, Rolland, Rollins, Romano, Romano, Romie, Rosińska,
  Röver, Rowan, Rüdiger, Ruggi, Ryan, Sakata, Sakosky, Salemi, Salit, Sammut,
  de~la Jordana, Sandberg, Sannibale, Santamaría, Santiago-Prieto, Santostasi,
  Saraf, Sassolas, Sathyaprakash, Sato, Satterthwaite, Saulson, Savage,
  Schilling, Schlamminger, Schnabel, Schofield, Schulz, Schutz, Schwinberg,
  Scott, Scott, Searle, Seifert, Sellers, Sengupta, Sentenac, Sergeev,
  Shaddock, Shaltev, Shapiro, Shawhan, Weerathunga, Shoemaker, Sibley, Siemens,
  Sigg, Singer, Singer, Sintes, Skelton, Slagmolen, Slutsky, Smith, Smith,
  Smith, Smith, Somiya, Sorazu, Soto, Speirits, Sperandio, Stefszky, Stein,
  Steinlechner, Steinlechner, Steplewski, Stochino, Stone, Strain, Strigin,
  Stroeer, Sturani, Stuver, Summerscales, Sung, Susmithan, Sutton, Swinkels,
  Szokoly, Tacca, Talukder, Tanner, Tarabrin, Taylor, Taylor, Thomas, Thorne,
  Thorne, Thrane, Thüring, Titsler, Tokmakov, Toncelli, Tonelli, Torre,
  Torres, Torrie, Tournefier, Travasso, Traylor, Trias, Tseng, Turner, Ugolini,
  Urbanek, Vahlbruch, Vaishnav, Vajente, Vallisneri, van~den Brand, Broeck,
  van~der Putten, van~der Sluys, van Veggel, Vass, Vasuth, Vaulin, Vavoulidis,
  Vecchio, Vedovato, Veitch, Veitch, Veltkamp, Verkindt, Vetrano, Viceré,
  Villar, Vinet, Vocca, Vorvick, Vyachanin, Waldman, Wallace, Wanner, Ward,
  Was, Wei, Weinert, Weinstein, Weiss, Wen, Wen, Wessels, West, Westphal,
  Wette, Whelan, Whitcomb, White, Whiting, Wilkinson, Willems, Williams,
  Williams, Willke, Winkelmann, Winkler, Wipf, Wiseman, Woan, Wooley, Worden,
  Yablon, Yakushin, Yamamoto, Yamamoto, Yang, Yeaton-Massey, Yoshida, Yu,
  Yvert, Zanolin, Zhang, Zhang, Zhao, Zotov, Zucker, , Zweizig, Collaboration,
  the Virgo~Collaboration, Aptekar, Boynton, Briggs, Cline, Connaughton,
  Frederiks, Gehrels, Goldsten, Golovin, van~der Horst, Hurley, Kaneko, von
  Kienlin, Kouveliotou, Krimm, Lin, Mitrofanov, Ohno, Pal'shin, Rau, Sanin,
  Tashiro, Terada, \& Yamaoka}]{2041-8205-734-2-L35}
---. 2011, \apjl, 734, L35

\bibitem[{Abadie {et~al.}(2012{\natexlab{a}})Abadie, {Abbott}, {Abbott},
  {Abbott}, {Abernathy}, {Accadia}, {Acernese}, {Adams}, {Adhikari}, \&
  et~al.}]{CBCLowLatency}
---. 2012{\natexlab{a}}, arXiv:1112.6005v4

\bibitem[{Abadie {et~al.}(2012{\natexlab{b}})Abadie, Abbott, {Abbott},
  {Abbott}, {Abernathy}, {Accadia}, {Acernese}, {Adams}, \&
  et~al.}]{loocup:2012}
---. 2012{\natexlab{b}}, \aap, 540, A124

\bibitem[{Abbott {et~al.}(2008)Abbott, Abbott, Adhikari, Ajith, Allen, Allen,
  Amin, Anderson, Anderson, Arain, Araya, Armandula, Armor, Aso, Aston,
  Aufmuth, Aulbert, Babak, Ballmer, Bantilan, Barish, Barker, Barker, Barr,
  Barriga, Barton, Bartos, Bastarrika, Bayer, Betzwieser, Beyersdorf, Bilenko,
  Billingsley, Biswas, Black, Blackburn, Blackburn, Blair, Bland, Bodiya,
  Bogue, Bork, Boschi, Bose, Brady, Braginsky, Brau, Brinkmann, Brooks, Brown,
  Brunet, Bullington, Buonanno, Burmeister, Byer, Cadonati, Cagnoli, Camp,
  Cannizzo, Cannon, Cao, Cardenas, Casebolt, Castaldi, Cepeda, Chalkley,
  Charlton, Chatterji, Chelkowski, Chen, Christensen, Clark, Clark, Cokelaer,
  Conte, Cook, Corbitt, Coyne, Creighton, Cumming, Cunningham, Cutler,
  Dalrymple, Danzmann, Davies, DeBra, Degallaix, Degree, Dergachev, Desai,
  DeSalvo, Dhurandhar, D\'{i}az, Dickson, Dietz, Donovan, Dooley, Doomes,
  Drever, Duke, Dumas, Dupuis, Dwyer, Echols, Effler, Ehrens, Espinoza, Etzel,
  Evans, Fairhurst, Fan, Fazi, Fehrmann, Fejer, Finn, Flasch, Fotopoulos,
  Freise, Frey, Fricke, Fritschel, Frolov, Fyffe, Garofoli, Gholami, Giaime,
  Giampanis, Giardina, Goda, Goetz, Goggin, Gonz\'alez, Gossler, Gouaty, Grant,
  Gras, Gray, Gray, Greenhalgh, Gretarsson, Grimaldi, Grosso, Grote, Grunewald,
  Guenther, Gustafson, Gustafson, Hage, Hallam, Hammer, Hanna, Hanson, Harms,
  Harry, Harstad, Hayama, Hayler, Heefner, Heng, Hennessy, Heptonstall,
  Hewitson, Hild, Hirose, Hoak, Hosken, Hough, Huttner, Ingram, Ito, Ivanov,
  Johnson, Johnson, Jones, Jones, Jones, Ju, Kalmus, Kalogera, Kamat, Kanner,
  Kasprzyk, Katsavounidis, Kawabe, Kawamura, Kawazoe, Kells, Keppel, Khalili,
  Khan, Khazanov, Kim, King, Kissel, Klimenko, Kokeyama, Kondrashov, Kopparapu,
  Kozak, Kozhevatov, Krishnan, Kwee, Lam, Landry, Lang, Lantz, Lazzarini, Lei,
  Leindecker, Leonhardt, Leonor, Libbrecht, Lin, Lindquist, Lockerbie, Lodhia,
  Lormand, Lu, Lubinski, Lucianetti, L\"uck, Machenschalk, MacInnis,
  Mageswaran, Mailand, Mandic, M\'arka, M\'arka, Markosyan, Markowitz, Maros,
  Martin, Martin, Marx, Mason, Matichard, Matone, Matzner, Mavalvala, McCarthy,
  McClelland, McGuire, McHugh, McIntyre, McIvor, McKechan, McKenzie, Meier,
  Melissinos, Mendell, Mercer, Meshkov, Messenger, Meyers, Miller, Minelli,
  Mitra, Mitrofanov, Mitselmakher, Mittleman, Miyakawa, Moe, Mohanty, Moreno,
  Mossavi, MowLowry, Mueller, Mukherjee, Mukhopadhyay, M\"uller-Ebhardt, Munch,
  Murray, Myers, Myers, Nash, Nelson, Newton, Nishizawa, Numata, O'Dell, Ogin,
  O'Reilly, O'Shaughnessy, Ottaway, Ottens, Overmier, Owen, Pan, Pankow, Papa,
  Parameshwaraiah, Patel, Pedraza, Penn, Perreca, Petrie, Pinto, Pitkin,
  Pletsch, Plissi, Postiglione, Principe, Prix, Quetschke, Raab, Rabeling,
  Radkins, Rainer, Rakhmanov, Ramsunder, Rehbein, Reid, Reitze, Riesen, Riles,
  Rivera, Robertson, Robinson, Robinson, Roddy, Rodriguez, Rogan, Rollins,
  Romano, Romie, Route, Rowan, R\"udiger, Ruet, Russell, Ryan, Sakata, Samidi,
  de~la Jordana, Sandberg, Sannibale, Saraf, Sarin, Sathyaprakash, Sato,
  Saulson, Savage, Savov, Schediwy, Schilling, Schnabel, Schofield, Schutz,
  Schwinberg, Scott, Searle, Sears, Seifert, Sellers, Sengupta, Shawhan,
  Shoemaker, Sibley, Siemens, Sigg, Sinha, Sintes, Slagmolen, Slutsky, Smith,
  Smith, Smith, Somiya, Sorazu, Stein, Stochino, Stone, Strain, Strom, Stuver,
  Summerscales, Sun, Sung, Sutton, Takahashi, Tanner, Taylor, Taylor, Thacker,
  Thorne, Thorne, Th\"uring, Tokmakov, Torres, Torrie, Traylor, Trias, Tyler,
  Ugolini, Ulmen, Urbanek, Vahlbruch, Van Den~Broeck, van~der Sluys, Vass,
  Vaulin, Vecchio, Veitch, Veitch, Villar, Vorvick, Vyachanin, Waldman,
  Wallace, Ward, Ward, Weinert, Weinstein, Weiss, Wen, Wette, Whelan, Whitcomb,
  Whiting, Wilkinson, Willems, Williams, Williams, Willke, Wilmut, Winkler,
  Wipf, Wiseman, Woan, Wooley, Worden, Wu, Yakushin, Yamamoto, Yan, Yoshida,
  Zanolin, Zhang, Zhang, Zhao, Zotov, Zucker, Zweizig, Barthelmy, Gehrels,
  Hurley, \& Palmer}]{PhysRevLett.101.211102}
Abbott, B., {et~al.} 2008, \prl, 101, 211102

\bibitem[{Abbott {et~al.}(2009)Abbott, Abbott, Adhikari, Ajith, Allen, Allen,
  Amin, Anderson, Anderson, Arain, Araya, Armandula, Armor, Aso, Aston,
  Aufmuth, Aulbert, Babak, Baker, Ballmer, Barker, Barker, Barr, Barriga,
  Barsotti, Barton, Bartos, Bassiri, Bastarrika, Behnke, Benacquista,
  Betzwieser, Beyersdorf, Bilenko, Billingsley, Biswas, Black, Blackburn,
  Blackburn, Blair, Bland, Bodiya, Bogue, Bork, Boschi, Bose, Brady, Braginsky,
  Brau, Bridges, Brinkmann, Brooks, Brown, Brummit, Brunet, Bullington,
  Buonanno, Burmeister, Byer, Cadonati, Camp, Cannizzo, Cannon, Cao, Cardenas,
  Caride, Castaldi, Caudill, Cavaglià, Cepeda, Chalermsongsak, Chalkley,
  Charlton, Chatterji, Chelkowski, Chen, Christensen, Chung, Clark, Clark,
  Clayton, Cokelaer, Colacino, Conte, Cook, Corbitt, Cornish, Coward, Coyne,
  Creighton, Creighton, Cruise, Culter, Cumming, Cunningham, Danilishin,
  Danzmann, Daudert, Davies, Daw, DeBra, Degallaix, Dergachev, Desai, DeSalvo,
  Dhurandhar, Díaz, Dietz, Donovan, Dooley, Doomes, Drever, Dueck, Duke,
  Dumas, Dwyer, Echols, Edgar, Effler, Ehrens, Espinoza, Etzel, Evans, Evans,
  Fairhurst, Faltas, Fan, Fazi, Fehrmenn, Finn, Flasch, Foley, Forrest,
  Fotopoulos, Franzen, Frede, Frei, Frei, Freise, Frey, Fricke, Fritschel,
  Frolov, Fyffe, Galdi, Garofoli, Gholami, Giaime, Giampanis, Giardina, Goda,
  Goetz, Goggin, González, Gorodetsky, Goßler, Gouaty, Grant, Gras, Gray,
  Gray, Greenhalgh, Gretarsson, Grimaldi, Grosso, Grote, Grunewald, Guenther,
  Gustafson, Gustafson, Hage, Hallam, Hammer, Hammond, Hanna, Hanson, Harms,
  Harry, Harry, Harstad, Haughian, Hayama, Heefner, Heng, Heptonstall,
  Hewitson, Hild, Hirose, Hoak, Hodge, Holt, Hosken, Hough, Hoyland, Hughey,
  Huttner, Ingram, Isogai, Ito, Ivanov, Johnson, Johnson, Jones, Jones, Jones,
  Ju, Kalmus, Kalogera, Kandhasamy, Kanner, Kasprzyk, Katsavounidis, Kawabe,
  Kawamura, Kawazoe, Kells, Keppel, Khalaidovski, Khalili, Khan, Khazanov,
  King, Kissel, Klimenko, Kokeyama, Kondrashov, Kopparapu, Koranda, Kozak,
  Krishnan, Kumar, Kwee, Lam, Landry, Lantz, Lazzarini, Lei, Lei, Leindecker,
  Leonor, Li, Lin, Lindquist, Littenberg, Lockerbie, Lodhia, Longo, Lormand,
  Lu, Lubinski, Lucianetti, Lück, Machenschalk, MacInnis, Mageswaran, Mailand,
  Mandel, Mandic, Márka, Márka, Markosyan, Markowitz, Maros, Martin, Martin,
  Marx, Mason, Matichard, Matone, Matzner, Mavalvala, McCarthy, McClelland,
  McGuire, McHugh, McIntyre, McKechan, McKenzie, Mehmet, Melatos, Melissinos,
  Menéndez, Mendell, Mercer, Meshkov, Messenger, Meyer, Miller, Minelli, Mino,
  Mitrofanov, Mitselmakher, Mittleman, Miyakawa, Moe, Mohanty, Mohapatra,
  Moreno, Morioka, Mors, Mossavi, MowLowry, Mueller, Müller-Ebhardt, Muhammad,
  Mukherjee, Mukhopadhyay, Mullavey, Munch, Murray, Myers, Myers, Nash, Nelson,
  Newton, Nishizawa, Numata, O'Dell, O'Reilly, O'Shaughnessy, Ochsner, Ogin,
  Ottaway, Ottens, Overmier, Owen, Pan, Pankow, Papa, Parameshwaraiah, Patel,
  Pedraza, Penn, Perraca, Pierro, Pinto, Pitkin, Pletsch, Plissi, Postiglione,
  Principe, Prix, Prokhorov, Punken, Quetschke, Raab, Rabeling, Radkins,
  Raffai, Raics, Rainer, Rakhmanov, Raymond, Reed, Reed, Rehbein, Reid, Reitze,
  Riesen, Riles, Rivera, Roberts, Robertson, Robinson, Robinson, Roddy, Röver,
  Rollins, Romano, Romie, Rowan, Rüdiger, Russell, Ryan, Sakata, de~la
  Jordana, Sandberg, Sannibale, Santamaría, Saraf, Sarin, Sathyaprakash, Sato,
  Satterthwaite, Saulson, Savage, Savov, Scanlan, Schilling, Schnabel,
  Schofield, Schulz, Schutz, Schwinberg, Scott, Scott, Searle, Sears, Seifert,
  Sellers, Sengupta, Sergeev, Shapiro, Shawhan, Shoemaker, Sibley, Siemens,
  Sigg, Sinha, Sintes, Slagmolen, Slutsky, Smith, Smith, Smith, Somiya, Sorazu,
  Stein, Stein, Steplewski, Stochino, Stone, Strain, Strigin, Stroeer, Stuver,
  Summerscales, Sun, Sung, Sutton, Szokoly, Talukder, Tang, Tanner, Tarabrin,
  Taylor, Taylor, Thacker, Thorne, Thüring, Tokmakov, Torres, Torrie, Traylor,
  Trias, Ugolini, Ulmen, Urbanek, Vahlbruch, Vallisneri, Broeck, van~der Sluys,
  van Veggel, Vass, Vaulin, Vecchio, Veitch, Veitch, Veltkamp, Villar, Vorvick,
  Vyachanin, Waldman, Wallace, Ward, Weidner, Weinert, Weinstein, Weiss, Wen,
  Wen, Wette, Whelan, Whitcomb, Whiting, Wilkinson, Willems, Williams,
  Williams, Willke, Wilmut, Winkelmann, Winkler, Wipf, Wiseman, Woan, Wooley,
  Worden, Wu, Yakushin, Yamamoto, Yan, Yoshida, Zanolin, Zhang, Zhang, Zhao,
  Zotov, Zucker, zur Mühlen, \& Zweizig}]{iLIGO}
Abbott, B.~P., {et~al.} 2009, Rep. Prog. Phys., 72, 076901

\bibitem[{Acernese {et~al.}(2008)Acernese, Alshourbagy, Amico, Antonucci,
  Aoudia, Arun, Astone, Avino, Baggio, Ballardin, Barone, Barsotti, Barsuglia,
  Bauer, Bigotta, Birindelli, Bizouard, Boccara, Bondu, Bosi, Braccini,
  Bradaschia, Brillet, Brisson, Buskulic, Cagnoli, Calloni, Campagna,
  Carbognani, Cavalier, Cavalieri, Cella, Cesarini, Chassande-Mottin,
  Chatterji, Cleva, Coccia, Corda, Corsi, Cottone, Coulon, Cuoco, D'Antonio,
  Dari, Dattilo, Davier, Rosa, Prete, Fiore, Lieto, Emilio, Virgilio, Evans,
  Fafone, Ferrante, Fidecaro, Fiori, Flaminio, Fournier, Frasca, Frasconi,
  Gammaitoni, Garufi, Genin, Gennai, Giazotto, Granata, Greverie, Grosjean,
  Guidi, Hamdani, Hebri, Heitmann, Hello, Huet, Penna, Laval, Leroy, Letendre,
  Lopez, Lorenzini, Loriette, Losurdo, Mackowski, Majorana, Man, Mantovani,
  Marchesoni, Marion, Marque, Martelli, Masserot, Menzinger, Milano, Minenkov,
  Mohan, Moreau, Morgado, Mosca, Mours, Neri, Nocera, Pagliaroli, Palomba,
  Paoletti, Pardi, Pasqualetti, Passaquieti, Passuello, Piergiovanni, Pinard,
  Poggiani, Punturo, Puppo, Rabaste, Rapagnani, Regimbau, Remillieux, Ricci,
  Ricciardi, Rocchi, Rolland, Romano, Ruggi, Sentenac, Solimeno, Swinkels,
  Terenzi, Toncelli, Tonelli, Tournefier, Travasso, Vajente, van~den Brand,
  van~der Putten, Verkindt, Vetrano, Viceré, Vinet, Vocca, \& Yvert}]{iVirgo}
Acernese, F., {et~al.} 2008, Class. Quantum Grav., 25, 184001

\bibitem[{Akerlof {et~al.}(2003)}]{rotse:2003}
Akerlof, C.~W., {et~al.} 2003, \pasp, 115, 132

\bibitem[{{Babul} {et~al.}(1987){Babul}, {Paczynski}, \&
  {Spergel}}]{1987ApJ...316L..49B}
{Babul}, A., {Paczynski}, B., \& {Spergel}, D. 1987, \apjl, 316, L49

\bibitem[{Baltay {et~al.}(2007)}]{quest:2007}
Baltay, C., {et~al.} 2007, \pasp, 119, 1278

\bibitem[{{Barron} {et~al.}(2011){Barron}, {Kaplan}, {Bangert}, {Bartlett},
  {Puatua}, {Harris}, \& {Barrett}}]{NOVAS}
{Barron}, E.~G., {Kaplan}, G.~H., {Bangert}, J., {Bartlett}, J.~L., {Puatua},
  W., {Harris}, W., \& {Barrett}, P. 2011, in \baas, Vol.~43, 217th Meeting of
  the American Astronomical Society, \#344.14

\bibitem[{Belczynski {et~al.}(2006)Belczynski, Perna, Bulik, Kalogera, Ivanova,
  \& Lamb}]{0004-637X-648-2-1110}
Belczynski, K., Perna, R., Bulik, T., Kalogera, V., Ivanova, N., \& Lamb, D.~Q.
  2006, \apj, 648, 1110

\bibitem[{Bennett {et~al.}(2010)Bennett, van Eysden, \& Melatos}]{MNR:MNR17416}
Bennett, M.~F., van Eysden, C.~A., \& Melatos, A. 2010, \mnras, 409, 1705

\bibitem[{Berezinsky {et~al.}(2001)Berezinsky, Hnatyk, \&
  Vilenkin}]{PhysRevD.64.043004}
Berezinsky, V., Hnatyk, B., \& Vilenkin, A. 2001, \prd, 64, 043004

\bibitem[{Bo\"{e}r {et~al.}(2003)}]{tarot:2003}
Bo\"{e}r, M., {et~al.} 2003, ESO Messenger, 113, 45

\bibitem[{Cannon {et~al.}(2012)Cannon, Cariou, Chapman, Crispin-Ortuzar,
  Fotopoulos, Frei, Hanna, Kara, Keppel, Liao, Privitera, Searle, Singer, \&
  Weinstein}]{lloid:2012}
Cannon, K., {et~al.} 2012, \apj, 748, 136

\bibitem[{Christensen \& Meyer(1998)}]{PhysRevD.58.082001}
Christensen, N., \& Meyer, R. 1998, \prd, 58, 082001

\bibitem[{Corsi \& Owen(2011)}]{PhysRevD.83.104014}
Corsi, A., \& Owen, B.~J. 2011, Phys. Rev. D, 83, 104014

\bibitem[{Coward {et~al.}(2010)}]{zadko:2010}
Coward, D.~M., {et~al.} 2010, \pasa, 27

\bibitem[{Damour \& Vilenkin(2000)}]{cscprl}
Damour, T., \& Vilenkin, A. 2000, \prl, 85, 3761

\bibitem[{Edmonds(1957)}]{edmonds1996angular}
Edmonds, A. 1957, Angular Momentum in Quantum Mechanics (Princeton University
  Press)

\bibitem[{Fairhurst(2009)}]{fairhurst:2009}
Fairhurst, S. 2009, New J. Phys., 11, 123006

\bibitem[{Fairhurst(2011)}]{fairhurst:2011}
---. 2011, Class. Quantum Grav., 28, 105021

\bibitem[{Gomboc {et~al.}(2004)}]{liverpool:2004}
Gomboc, A., {et~al.} 2004, Nucl. Phys. B Proc. Supp., 132, 312

\bibitem[{G\'{o}rski {et~al.}(2005)G\'{o}rski, Hivon, Banday, Wandelt, Hansen,
  Reinecke, \& Bartelmann}]{healpix}
G\'{o}rski, K.~M., Hivon, E., Banday, A.~J., Wandelt, B.~D., Hansen, F.~K.,
  Reinecke, M., \& Bartelmann, M. 2005, \apj, 622, 759

\bibitem[{{Guidorzi} {et~al.}(2006){Guidorzi}, {Monfardini}, {Gomboc},
  {Mottram}, {Mundell}, {Steele}, {Carter}, {Bode}, {Smith}, {Fraser},
  {Burgdorf}, \& {Newsam}}]{liverpool:2006}
{Guidorzi}, C., {et~al.} 2006, \pasp, 118, 288

\bibitem[{Harry \& the LIGO Scientific~Collaboration(2010)}]{aLIGO}
Harry, G.~M., \& the LIGO Scientific~Collaboration. 2010, Class. Quantum Grav.,
  27, 084006

\bibitem[{Ingber(1989)}]{Ingber1989967}
Ingber, L. 1989, Mathematical and Computer Modelling, 12, 967

\bibitem[{Ioka(2001)}]{MNR:MNR4756}
Ioka, K. 2001, \mnras, 327, 639

\bibitem[{Kanner {et~al.}(2008)Kanner, Huard, M\'arka, Murphy, Piscionere,
  Reed, \& Shawhan}]{loocup:2008}
Kanner, J., Huard, T.~L., M\'arka, S., Murphy, D.~C., Piscionere, J., Reed, M.,
  \& Shawhan, P. 2008, Class. Quantum Grav., 25, 184034

\bibitem[{Keller {et~al.}(2007)}]{skymapper:2007}
Keller, S.~C., {et~al.} 2007, \pasa, 24, 1

\bibitem[{{Klimenko} {et~al.}(2011){Klimenko}, {Vedovato}, {Drago}, {Mazzolo},
  {Mitselmakher}, {Pankow}, {Prodi}, {Re}, {Salemi}, \&
  {Yakushin}}]{klimenko:2011}
{Klimenko}, S., {et~al.} 2011, \prd, 83, 102001

\bibitem[{Klotz {et~al.}(2008)}]{tarot:2008}
Klotz, A., {et~al.} 2008, \pasp, 120, 1298

\bibitem[{Kocsis {et~al.}(2008)Kocsis, Haiman, \& Menou}]{kocsis:2008}
Kocsis, B., Haiman, Z., \& Menou, K. 2008, \apj, 684, 870

\bibitem[{Kuroda(2010)}]{LCGT}
Kuroda, K. 2010, Class. Quantum Grav., 27, 084004

\bibitem[{Law {et~al.}(2009)}]{ptf:2009}
Law, N.~M., {et~al.} 2009, \pasp, 121, 1395

\bibitem[{Lee \& Ramirez-Ruiz(2007{\natexlab{a}})}]{1367-2630-9-1-017}
Lee, W.~H., \& Ramirez-Ruiz, E. 2007{\natexlab{a}}, New J. Phys., 9, 17

\bibitem[{Lee \& Ramirez-Ruiz(2007{\natexlab{b}})}]{lee}
---. 2007{\natexlab{b}}, New J. Phys., 7

\bibitem[{Majcher {et~al.}(2011)}]{pi:2011}
Majcher, A., {et~al.} 2011, \procspie, 8008

\bibitem[{Ma{\l}ek {et~al.}(2010)}]{pi:2010}
Ma{\l}ek, K., {et~al.} 2010, Adv. Astron., 2010, 9

\bibitem[{{Manzotti} \& {Dietz}(2012)}]{dietz:2012}
{Manzotti}, A., \& {Dietz}, A. 2012, arXiv:1202.4031

\bibitem[{McEwen \& Wiaux(2011)}]{mcewen}
McEwen, J., \& Wiaux, Y. 2011, IEEE Trans. Signal Process., 59, 5876

\bibitem[{M\'{e}sz\'{a}ros(2006)}]{0034-4885-69-8-R01}
M\'{e}sz\'{a}ros, P. 2006, Rep. Prog. Phys., 69, 2259

\bibitem[{Metzger \& Berger(2012)}]{MetzgerBerger:2012}
Metzger, B.~D., \& Berger, E. 2012, \apj, 746, 48

\bibitem[{{Nakar}(2007)}]{2007PhR...442..166N}
{Nakar}, E. 2007, \physrep, 442, 166

\bibitem[{Nissanke {et~al.}(2010)Nissanke, Holz, Hughes, Dalal, \&
  Sievers}]{Nissanke:2010}
Nissanke, S., Holz, D.~E., Hughes, S.~A., Dalal, N., \& Sievers, J.~L. 2010,
  \apj, 725, 496

\bibitem[{Nissanke {et~al.}(2011)Nissanke, Sievers, Dalal, \&
  Holz}]{Nissanke:2011}
Nissanke, S., Sievers, J., Dalal, N., \& Holz, D. 2011, \apj, 739, 99

\bibitem[{Ott(2009)}]{ott}
Ott, C.~D. 2009, Class. Quantum Grav., 26, 063001

\bibitem[{Press {et~al.}(2007)Press, Teukolsky, Vetterling, \&
  Flannery}]{numerical-recipes-chapter-13}
Press, W.~H., Teukolsky, S.~A., Vetterling, W.~T., \& Flannery, B.~P. 2007,
  Numerical Recipes, 3rd edn. (Cambridge University Press), 641--647

\bibitem[{Raymond {et~al.}(2009)Raymond, van~der Sluys, Mandel, Kalogera,
  Röver, \& Christensen}]{Raymound:2009p114007}
Raymond, V., van~der Sluys, M.~V., Mandel, I., Kalogera, V., Röver, C., \&
  Christensen, N. 2009, Class. Quantum Grav., 26, 114007

\bibitem[{Risbo(1996)}]{risbo}
Risbo, T. 1996, J. Geodesy, 70, 383, 10.1007/BF01090814

\bibitem[{Sakurai \& Tuan(1994)}]{sakurai1994modern}
Sakurai, J., \& Tuan, S. 1994, Modern quantum mechanics (Addison-Wesley Pub.
  Co.)

\bibitem[{Sathyaprakash \& Schutz(2009)}]{lrr-2009-2}
Sathyaprakash, B., \& Schutz, B.~F. 2009, Living Rev. Relativ., 12

\bibitem[{Schutz(2001)}]{schutz2001}
Schutz, B.~F. 2001, in Proceedings of the {MPA/ESO/MPE/USM} Joint Astronomy
  Conference, ed. M.~Gilfanov, R.~Sunyaev, \& E.~Churazov, {ESO} Astrophysics
  Symposia, Garching, Germany

\bibitem[{Schutz(2011)}]{0264-9381-28-12-125023}
Schutz, B.~F. 2011, Class. Quantum Grav., 28, 125023

\bibitem[{Searle {et~al.}(2009)Searle, Sutton, \& Tinto}]{Searle:2009}
Searle, A.~C., Sutton, P.~J., \& Tinto, M. 2009, Class. Quantum Grav., 26,
  155017

\bibitem[{Searle {et~al.}(2008)Searle, Sutton, Tinto, \& Woan}]{Searle:2008}
Searle, A.~C., Sutton, P.~J., Tinto, M., \& Woan, G. 2008, Class. Quantum
  Grav., 25, 114038

\bibitem[{Siemens {et~al.}(2006)Siemens, Creighton, Maor, Majumder, Cannon, \&
  Read}]{cscgw}
Siemens, X., Creighton, J., Maor, I., Majumder, S.~R., Cannon, K., \& Read, J.
  2006, \prd, 73, 105001

\bibitem[{Sylvestre(2003)}]{sylvestre}
Sylvestre, J. 2003, \apj, 591, 1152

\bibitem[{Troja {et~al.}(2008)Troja, King, O'Brien, Lyons, \&
  Cusumano}]{MNL2:MNL20421}
Troja, E., King, A.~R., O'Brien, P.~T., Lyons, N., \& Cusumano, G. 2008,
  \mnras, 385, L10

\bibitem[{{U.S. Nautical Almanac Office}(2010)}]{u2010astronomical}
{U.S. Nautical Almanac Office}. 2010, Astronomical Almanac: 2011, ASTRONOMICAL
  ALMANAC FOR THE YEAR (Bernan Assoc)

\bibitem[{Vachaspati(2008)}]{cscem}
Vachaspati, T. 2008, \prl, 101, 141301

\bibitem[{van~der Sluys {et~al.}(2008)van~der Sluys, Raymond, Mandel, Röver,
  Christensen, Kalogera, Meyer, \& Vecchio}]{0264-9381-25-18-184011}
van~der Sluys, M., Raymond, V., Mandel, I., Röver, C., Christensen, N.,
  Kalogera, V., Meyer, R., \& Vecchio, A. 2008, Class. Quantum Grav., 25,
  184011

\bibitem[{Veitch \& Vecchio(2010)}]{PhysRevD.81.062003}
Veitch, J., \& Vecchio, A. 2010, \prd, 81, 062003

\bibitem[{{Virgo Collaboration}(2009)}]{AdVirgo}
{Virgo Collaboration}. 2009, VIR-027A-09,
  \url{https://tds.ego-gw.it/ql/?c=6589}

\bibitem[{Wandelt \& G\'orski(2001)}]{wandelt}
Wandelt, B.~D., \& G\'orski, K.~M. 2001, \prd, 63, 123002

\bibitem[{Wang {et~al.}(2011)Wang, Fan, \& Wei}]{PhysRevLett.106.259001}
Wang, Y., Fan, Y.-Z., \& Wei, D.-M. 2011, \prl, 106, 259001

\bibitem[{Wen \& Chen(2010)}]{Wen:2010}
Wen, L., \& Chen, Y. 2010, Phys. Rev. D, 81, 082001

\bibitem[{White(2012)}]{white:2012}
White, D. 2012, personal communication

\end{thebibliography}

\end{document}